\documentclass[12pt,english]{article}
\usepackage{tgtermes}
\usepackage{helvet}
\usepackage[lf]{FiraMono}

\usepackage[T1]{fontenc}
\usepackage[latin9]{inputenc}
\usepackage[table]{xcolor}
\usepackage{colortbl}
\usepackage{array}
\usepackage{verbatim}
\usepackage{mathtools}
\usepackage{multirow}
\usepackage{varwidth}
\usepackage{amsmath}
\usepackage{amsthm}
\usepackage{amssymb}
\usepackage[a4paper]{geometry}
\usepackage[authoryear]{natbib}

\makeatletter

\providecommand{\tabularnewline}{\\}
\newenvironment{cellvarwidth}[1][t]
    {\begin{varwidth}[#1]{\linewidth}}
    {\@finalstrut\@arstrutbox\end{varwidth}}

\theoremstyle{plain}
\newtheorem{theorem}{Theorem}[section]
\newtheorem{lemma}{Lemma}[section]

\theoremstyle{definition}
\newtheorem{assumption}{Assumption}[section]

\theoremstyle{remark}
\newtheorem{remark}{Remark}[section]

\usepackage[section]{placeins}

\usepackage{microtype}

\usepackage{pdfpages}

\makeatother

\usepackage{babel}
\begin{document}
\title{Curvature-Calibrated Quasi-Bayesian Updating for Moment-Restricted
Models}
\author{Masahiro Tanaka\thanks{Faculty of Economics, Fukuoka University, Fukuoka, Japan. Address:
8-19-1, Nanakuma, Jonan, Fukuoka, Japan 814-0180. E-mail: m.tanaka.tt@fukuoka-u.ac.jp.}}
\maketitle
\begin{abstract}
Quadratic moment-based quasi-posteriors depend on a weighting matrix
governing both the contributions of moment conditions and the scale
of posterior uncertainty. We propose curvature-calibrated quasi-Bayesian
(CCQB) updating, which iteratively evaluates the inverse of the estimated
moment covariance (or long-run covariance) at the center of the resulting
fixed-weight quasi-posterior. A two-part stopping rule monitors changes
in the center and Monte Carlo error in estimating it. Under local
regularity conditions, the ideal calibration map is contractive, and
its fixed point is root-$N$ consistent. The calibrated quasi-posterior
and the sampling distribution of its center have the inverse Godambe
information matrix as their root-$N$-scale covariance. Simulations
show substantial gains over identity weighting. Further iteration
has little effect in linear designs but improves calibration for some
parameters in nonlinear settings, although nominal coverage is not
uniformly achieved. An application to longitudinal childhood wheeze
data illustrates the method under within-subject dependence.
\end{abstract}

\section{Introduction}

\label{sec:intro} 

Moment restrictions provide a useful basis for Bayesian-style inference
when a complete likelihood is unavailable, difficult to evaluate,
or deliberately left unspecified. A growing literature constructs
quasi-posteriors by combining a prior distribution with an empirical
criterion derived from moment conditions \citep{Kim2002,Chernozhukov2003,Yin2009,Liao2011,Li2016}.
These methods retain familiar Bayesian computational and inferential
tools, including Markov chain Monte Carlo (MCMC), while allowing the
empirical information to be specified through moment conditions rather
than a full probability model. This flexibility is particularly attractive
in settings with complex dependence structures, latent variables,
or other features for which specifying and computing a full likelihood
can be burdensome.

An important issue arises, however, when a quadratic moment criterion
serves as a surrogate for a log likelihood. In an ordinary likelihood-based
posterior, the likelihood determines the local information scale on
which the evidence in the data is combined with the prior. A moment
criterion has no analogous intrinsic scale. For a quasi-posterior
based on a quadratic form in the sample moments, the weighting matrix
determines both the overall magnitude and the directional curvature
of the empirical criterion. Its relative structure controls the weighting
of different moment conditions, while its overall magnitude acts as
a learning rate. Consequently, even with the same moments and prior,
different weighting matrices can yield different quasi-posteriors,
including different degrees of prior influence in finite samples.
Although the moment restrictions do not determine a uniquely correct
finite-sample balance between the prior and the empirical criterion,
they provide a basis for choosing its local information scale through
the sampling covariance of the moments.

This distinction has no direct counterpart in classical point estimation
by the generalized method of moments (GMM) \citep{Hansen1982}. In
particular, multiplying a GMM weighting matrix by a positive scalar
leaves the minimizer of the criterion unchanged. For quasi-Bayesian
updating, the same rescaling changes the dispersion of the quasi-posterior
and, in finite samples, can also change its center through its interaction
with the prior. More generally, an arbitrary weighting matrix can
cause the covariance represented by the quasi-posterior to differ
from the repeated-sampling covariance of its center. Related work
by \citet{Kleijn2012} shows that, under model misspecification, a
posterior may satisfy a Bernstein--von Mises approximation while
its local covariance fails to reproduce the repeated-sampling variability
of its center. Thus, posterior curvature and frequentist uncertainty
need not coincide once the likelihood-based information identity is
lost. This observation motivates choosing the weighting matrix both
to support efficient point estimation and to place the empirical component
of the quasi-posterior on an appropriate local information scale.

This paper proposes curvature-calibrated quasi-Bayesian (CCQB) updating
for this purpose. The basic idea is to use the inverse of the local
covariance, or long-run covariance, of the moment conditions as the
weighting matrix and to evaluate that covariance at a center determined
self-consistently by the resulting quasi-posterior. Starting from
a pilot quasi-posterior, CCQB alternates between estimating the covariance
of the moments at the current quasi-posterior center and simulating
a fixed-weight quasi-posterior using its inverse. A two-part stopping
rule separately assesses changes between successive centers, standardized
by a reference covariance matrix, and the Monte Carlo uncertainty
in their estimation. After the rule is met, a final quasi-posterior
is simulated using the calibrated weight. Thus, CCQB calibrates the
full local curvature of the empirical criterion rather than only its
scalar magnitude. At the same time, the weighting matrix remains fixed
within each MCMC run, avoiding the potentially substantial cost of
evaluating a parameter-dependent covariance matrix and its determinant
at every MCMC iteration.

The calibration has a direct information-geometric interpretation.
Let $\boldsymbol{G}_{0}$ denote the population derivative of the
moments and $\boldsymbol{C}_{0}$ their long-run covariance at the
true parameter. For a generic fixed weighting matrix, the first-order
sampling covariance of the quasi-posterior center need not agree with
the covariance represented by the quasi-posterior itself. With the
covariance weight $\boldsymbol{C}^{-1}_{0}$, the asymptotic covariance
matrices of the corresponding root-$N$ scaled quantities both reduce
to $(\boldsymbol{G}^{\top}_{0}\boldsymbol{C}^{-1}_{0}\boldsymbol{G}_{0})^{-1}$,
the inverse of the Godambe information matrix associated with the
specified moments. CCQB estimates this locally appropriate weight
while resolving the dependence of the covariance estimate on the unknown
parameter through a fixed-point condition. In this sense, the procedure
calibrates the empirical curvature before the final quasi-posterior
is formed, so that the prior is combined with a data component having
an interpretable local information scale.

We establish several large-sample properties of this procedure. Under
local regularity conditions, the ideal CCQB calibration map is, with
probability approaching one, a contraction in a neighborhood of the
true parameter, with a contraction modulus of order $N^{-1/2}$. Its
fixed point is root-$N$ consistent, and the corresponding weighting
matrix converges to $\boldsymbol{C}^{-1}_{0}$. The calibrated quasi-posterior
satisfies a Bernstein--von Mises approximation with asymptotic covariance
equal to the inverse Godambe information matrix, and the fixed-point
center has the same first-order sampling covariance. The contraction
result also shows that only a small number of calibration updates
are needed asymptotically: from any initial value in the local contraction
region, two ideal updates are first-order equivalent to the fixed
point, while one update suffices when the initial center is consistent.
We additionally give conditions under which a pilot MCMC quasi-posterior
provides such an initial center and characterize the Monte Carlo accuracy
required for the computed calibration iterates.

A fixed-weight quasi-posterior based on a covariance estimate evaluated
at any consistent preliminary center already achieves first-order
covariance calibration. Self-consistent iteration is therefore not
required for asymptotic validity. Its contribution is instead to select
the covariance-evaluation point through the center of the resulting
quasi-posterior, with theoretical guarantees for local stability and
convergence. The potential finite-sample benefits of this additional
adjustment depend on the model and are assessed empirically rather
than assumed.

The simulation study separates the first covariance-weight update
from further iteration and compares both with a sample-oracle weight
and Bayesian exponentially tilted empirical likelihood (BETEL) \citep{Chib2018}.
The experiments use an overidentified linear instrumental-variables
design with eight moments, a nonlinear, exactly identified Poisson
design, and a heteroskedastic and serially dependent regression design.
For nominal $95\%$ intervals, covariance weighting produces most
of the improvement in the linear designs, with little additional effect
from iteration. In the Poisson design, iteration further improves
covariance calibration and coverage for some coefficients, although
notable undercoverage remains for others. BETEL typically yields shorter
intervals with lower coverage in smaller samples. The stopping rule
requires only a few updates in the reported designs. An application
to longitudinal childhood wheeze data illustrates the estimation of
 child-level moment covariance in the presence of within-child dependence
and informative priors.

The proposed framework is related to several strands of work. A complementary
strand of the literature develops Bayesian procedures for moment-restricted
models using likelihood-like constructions based on empirical likelihood
or exponential tilting, including Bayesian empirical likelihood \citep{Lazar2003},
Bayesian exponentially tilted empirical likelihood \citep{Schennach2005},
and the Bayesian semiparametric approach of \citet{Chib2018}. These
methods and CCQB share the goal of conducting Bayesian inference without
a fully specified parametric likelihood. CCQB differs in that it retains
a quadratic moment quasi-posterior and calibrates its weighting matrix,
and hence its local information scale. The two constructions need
not have identical higher-order or finite-sample behavior, as shown
in the simulation study.

CCQB is also related to calibration methods for generalized and Gibbs
posteriors. Generalized Bayesian procedures replace the negative log
likelihood by a loss function \citep{Zhang2006a,Zhang2006b,Jiang2008,Bissiri2016,Miller2021},
while several methods calibrate a scalar learning rate to control
the scale of the loss relative to the prior \citep{Gruenwald2017,Holmes2017,Lyddon2019,Syring2019}.
A scalar adjustment, however, cannot generally correct multidimensional
discrepancies in posterior curvature \citep{Miller2021,Tanaka2026a}. 

More recently, \citet{Frazier2026} construct a covariance-weighted
quadratic criterion from the gradient of a generic loss, yielding
asymptotically calibrated uncertainty without scalar tuning. Like
their approach, CCQB uses covariance weighting to align posterior
and repeated-sampling uncertainty. However, their estimating-equation
representation corresponds to a just-identified system, whereas CCQB
directly accommodates both just-identified and overidentified moment
conditions. CCQB reduces the per-evaluation computational burden by
fixing the covariance weight within each simulation run, at the cost
of a small number of calibration runs.

CCQB further differs from probability-calibration methods and post-simulation
sandwich corrections. Probability calibration targets agreement between
posterior credible levels and repeated-sampling coverage probabilities
\citep{Dawid1982}, whereas CCQB targets agreement between local quasi-posterior
curvature and the Godambe information implied by the moment restrictions.
Likewise, sandwich adjustments such as \citet{Mueller2013} and \citet{Shaby2014}
modify uncertainty after a posterior or quasi-posterior has been constructed,
whereas CCQB calibrates the empirical criterion before the final quasi-posterior
is simulated. Although CCQB also resembles iterated GMM \citep{Hansen1996}
in its use of an updated covariance-based weight, the overall scale
of that weight is irrelevant for classical GMM point estimation but
determines the balance between the empirical criterion and the prior
in quasi-Bayesian inference.

The remainder of the paper is organized as follows. Section~\ref{sec:method}
formulates the calibration problem, introduces the CCQB fixed-point
algorithm, and establishes its large-sample properties. Section~\ref{sec:simulation}
reports the simulation study. Section~\ref{sec:application} presents
the empirical application. The final section concludes.

\section{Method}

\label{sec:method}

\subsection{Moment quasi-posteriors and the calibration problem}

\label{subsec:setup}

Let $\mathcal{D}_{N}=(D_{1},\dots,D_{N})$ denote the sample, let
$\boldsymbol{\theta}\in\Theta\subset\mathbb{R}^{J}$ be the parameter
of interest, and let $\boldsymbol{m}(D_{i},\boldsymbol{\theta})\in\mathbb{R}^{K}$,
with $K\geq J$, be a vector of moment functions. For notational convenience,
we suppress the dependence on $D_{i}$ and write $\boldsymbol{m}_{i}(\boldsymbol{\theta})=\boldsymbol{m}(D_{i},\boldsymbol{\theta})$.
Define the sample mean of $\boldsymbol{m}_{i}(\boldsymbol{\theta})$
as 
\[
\bar{\boldsymbol{m}}_{N}(\boldsymbol{\theta})=\frac{1}{N}\sum^{N}_{i=1}\boldsymbol{m}_{i}(\boldsymbol{\theta}),
\]
and suppose that 
\[
\mathbb{E}\{\boldsymbol{m}_{i}(\boldsymbol{\theta}_{0})\}=\boldsymbol{0}_{K},
\]
where $\boldsymbol{\theta}_{0}$ is the unique zero of the map $\boldsymbol{\theta}\mapsto\mathbb{E}\{\boldsymbol{m}_{i}(\boldsymbol{\theta})\}$
in a neighborhood of $\boldsymbol{\theta}_{0}$, and $\boldsymbol{0}_{K}$
denotes the $K$-dimensional zero vector.

Let $\mathbb{S}^{K}$ denote the vector space of symmetric $K\times K$
matrices, and let $\mathbb{S}^{K}_{++}$ denote its positive-definite
cone. All topological notions for subsets of $\mathbb{S}^{K}$ are
understood relative to $\mathbb{S}^{K}$. Throughout, $\lVert\cdot\rVert$
denotes the Euclidean norm for vectors and the induced spectral norm
for matrices. For a weakly stationary sequence, let
\[
\boldsymbol{\Gamma}_{h}(\boldsymbol{\theta})=\operatorname{Cov}\{\boldsymbol{m}_{i}(\boldsymbol{\theta}),\boldsymbol{m}_{i-h}(\boldsymbol{\theta})\}
\]
denote the lag-$h$ autocovariance matrix, and assume that these matrices
are absolutely summable. Define the population long-run covariance
matrix by
\[
\boldsymbol{C}(\boldsymbol{\theta})=\sum_{h\in\mathbb{Z}}\boldsymbol{\Gamma}_{h}(\boldsymbol{\theta}).
\]
We assume below that $\boldsymbol{C}(\boldsymbol{\theta})$ is positive
definite in a neighborhood of $\boldsymbol{\theta}_{0}$. Under independence,
the long-run covariance reduces to the population covariance matrix:
\[
\boldsymbol{C}(\boldsymbol{\theta})=\operatorname{Var}\{\boldsymbol{m}_{i}(\boldsymbol{\theta})\}.
\]
Let $\boldsymbol{C}_{0}=\boldsymbol{C}(\boldsymbol{\theta}_{0})$
and $\widehat{\boldsymbol{C}}_{N}(\boldsymbol{\theta})$ denote an
estimator of $\boldsymbol{C}(\boldsymbol{\theta})$. Define the weighting
matrix by 
\[
\boldsymbol{W}_{N}(\boldsymbol{\theta})=\widehat{\boldsymbol{C}}_{N}(\boldsymbol{\theta})^{-1}.
\]

Motivated by the central limit theorem 
\[
\sqrt{N}\bar{\boldsymbol{m}}_{N}(\boldsymbol{\theta}_{0})\rightsquigarrow\mathcal{N}(\boldsymbol{0}_{K},\boldsymbol{C}_{0}),
\]
we use a Gaussian working likelihood for the sample moments, yielding
the quasi-posterior: 
\begin{equation}
\Pi^{\boldsymbol{W}_{N}(\cdot)}(d\boldsymbol{\theta}\mid\mathcal{D}_{N})\propto\det\{\boldsymbol{W}_{N}(\boldsymbol{\theta})\}^{1/2}\exp\left\{ -\frac{N}{2}\bar{\boldsymbol{m}}_{N}(\boldsymbol{\theta})^{\top}\boldsymbol{W}_{N}(\boldsymbol{\theta})\bar{\boldsymbol{m}}_{N}(\boldsymbol{\theta})\right\} \pi(\boldsymbol{\theta})\,d\boldsymbol{\theta},\label{eq:quasi-posterior}
\end{equation}
where $\pi(\boldsymbol{\theta})$ denotes the prior density. Quasi-posterior
draws of $\boldsymbol{\theta}$ are simulated using standard Bayesian
computational methods. We use MCMC terminology below for concreteness,
although alternatives such as slice sampling \citep{Neal2003} and
sequential Monte Carlo sampling \citep{DelMoral2006} are also feasible.
Evaluating $\boldsymbol{W}_{N}(\boldsymbol{\theta})$ and its determinant
at every MCMC iteration can, however, be computationally expensive
and numerically unstable \citep{Yin2011,Tanaka2025,Tanaka2026b}.
A common alternative is to fix a positive definite matrix $\boldsymbol{W}$
and simulate from 
\begin{equation}
\Pi^{\boldsymbol{W}}(d\boldsymbol{\theta}\mid\mathcal{D}_{N})\propto\exp\left\{ -\frac{N}{2}\bar{\boldsymbol{m}}_{N}(\boldsymbol{\theta})^{\top}\boldsymbol{W}\bar{\boldsymbol{m}}_{N}(\boldsymbol{\theta})\right\} \pi(\boldsymbol{\theta})\,d\boldsymbol{\theta}.\label{eq:fixed-W-quasi-posterior}
\end{equation}

Unlike an ordinary likelihood, the quadratic moment criterion has
no intrinsic information scale. The matrix $\boldsymbol{W}$ controls
the local information metric, whereas its relative eigenstructure
determines the relative weighting of the moment conditions. Thus,
different choices of $\boldsymbol{W}$ can produce different quasi-posteriors
even when the moments and the prior are unchanged.

To clarify the implications, consider the asymptotic behavior of the
moment-based quasi-posterior. Let 
\[
\boldsymbol{G}_{0}=\mathbb{E}\left\{ \frac{\partial\boldsymbol{m}_{i}(\boldsymbol{\theta}_{0})}{\partial\boldsymbol{\theta}^{\top}}\right\} \in\mathbb{R}^{K\times J}.
\]
Assume that $\boldsymbol{G}_{0}$ has full column rank. For a fixed
$\boldsymbol{W}\in\mathbb{S}^{K}_{++}$, define the quasi-posterior
mean 
\[
\boldsymbol{\mu}_{N}(\boldsymbol{W})=\mathbb{E}_{\Pi^{\boldsymbol{W}}(\cdot\mid\mathcal{D}_{N})}(\boldsymbol{\theta}).
\]
Suppose that the corresponding fixed-weight quasi-posterior satisfies
the usual pointwise regularity conditions, including a local quadratic
expansion, a Bernstein--von Mises approximation, and posterior moment
conditions sufficient to transfer the local Gaussian approximation
to its first two moments \citep{Chernozhukov2003,Hong2021}. Then
\[
\sqrt{N}\{\boldsymbol{\mu}_{N}(\boldsymbol{W})-\boldsymbol{\theta}_{0}\}=-\boldsymbol{A}(\boldsymbol{W})\sqrt{N}\bar{\boldsymbol{m}}_{N}(\boldsymbol{\theta}_{0})+o_{p}(1),
\]
where 
\[
\boldsymbol{A}(\boldsymbol{W})=\left(\boldsymbol{G}^{\top}_{0}\boldsymbol{W}\boldsymbol{G}_{0}\right)^{-1}\boldsymbol{G}^{\top}_{0}\boldsymbol{W}.
\]
Consequently, if 
\[
\sqrt{N}\bar{\boldsymbol{m}}_{N}(\boldsymbol{\theta}_{0})\rightsquigarrow\mathcal{N}(\boldsymbol{0}_{K},\boldsymbol{C}_{0}),
\]
then 
\[
\sqrt{N}\{\boldsymbol{\mu}_{N}(\boldsymbol{W})-\boldsymbol{\theta}_{0}\}\rightsquigarrow\mathcal{N}\bigl(\boldsymbol{0}_{J},\boldsymbol{V}(\boldsymbol{W})\bigr),
\]
with 
\[
\boldsymbol{V}(\boldsymbol{W})=\boldsymbol{A}(\boldsymbol{W})\boldsymbol{C}_{0}\boldsymbol{A}(\boldsymbol{W})^{\top}.
\]
By contrast, under the same pointwise conditions, the conditional
covariance of the fixed-weight quasi-posterior satisfies 
\[
N\operatorname{Var}_{\Pi^{\boldsymbol{W}}(\cdot\mid\mathcal{D}_{N})}(\boldsymbol{\theta})\xrightarrow{p}\left(\boldsymbol{G}^{\top}_{0}\boldsymbol{W}\boldsymbol{G}_{0}\right)^{-1}.
\]
For later use, define 
\[
\begin{aligned}\boldsymbol{W}_{0} & =\boldsymbol{C}^{-1}_{0}, & \boldsymbol{A}_{0} & =\boldsymbol{A}(\boldsymbol{W}_{0}),\\
\boldsymbol{V}_{0} & =\boldsymbol{A}_{0}\boldsymbol{C}_{0}\boldsymbol{A}^{\top}_{0}=\left(\boldsymbol{G}^{\top}_{0}\boldsymbol{C}^{-1}_{0}\boldsymbol{G}_{0}\right)^{-1}, & \boldsymbol{\xi}_{N} & =\sqrt{N}\bar{\boldsymbol{m}}_{N}(\boldsymbol{\theta}_{0}).
\end{aligned}
\]

Thus, the frequentist asymptotic covariance of the quasi-posterior
mean need not agree with the covariance represented by the quasi-posterior.
A natural choice that makes the two covariance matrices coincide is
$\boldsymbol{W}=\boldsymbol{C}^{-1}_{0}$. In that case, both are
equal to $\left(\boldsymbol{G}^{\top}_{0}\boldsymbol{C}^{-1}_{0}\boldsymbol{G}_{0}\right)^{-1}$.
Although every positive scalar multiple of $\boldsymbol{C}^{-1}_{0}$
yields the same efficient GMM point estimator, that scalar changes
the quasi-posterior dispersion and therefore cannot be left unspecified
here.

Here, first-order covariance calibration refers to agreement between
the asymptotic covariance represented by the quasi-posterior and the
repeated-sampling covariance of its center. Together with the corresponding
Bernstein--von Mises approximation and central limit theorem, this
agreement implies asymptotically correct frequentist coverage for
regular credible intervals and ellipsoids. It does not imply exact
finite-sample coverage.

Before introducing CCQB, it is useful to separate calibration of the
quasi-posterior itself from a post-simulation covariance correction.
A sandwich-adjusted covariance estimator can be applied after simulation.
For a generic evaluation point $\boldsymbol{\theta}^{\dagger}\in\Theta$,
define 
\[
\widehat{\boldsymbol{G}}_{N}\left(\boldsymbol{\theta}^{\dagger}\right)=\left.\frac{1}{N}\sum^{N}_{i=1}\frac{\partial\boldsymbol{m}_{i}(\boldsymbol{\theta})}{\partial\boldsymbol{\theta}^{\top}}\right|_{\boldsymbol{\theta}=\boldsymbol{\theta}^{\dagger}}.
\]
Given retained draws $\boldsymbol{\theta}_{1},\dots,\boldsymbol{\theta}_{T}$
from $\Pi^{\boldsymbol{W}}(d\boldsymbol{\theta}\mid\mathcal{D}_{N})$,
let 
\[
\widetilde{\boldsymbol{\mu}}_{N,T}(\boldsymbol{W})=\frac{1}{T}\sum^{T}_{t=1}\boldsymbol{\theta}_{t}
\]
denote the MCMC estimate of the target quasi-posterior mean $\boldsymbol{\mu}_{N}(\boldsymbol{W})$
and set $\boldsymbol{\theta}^{\dagger}=\widetilde{\boldsymbol{\mu}}_{N,T}(\boldsymbol{W})$.
Define the sample analogue of $\boldsymbol{A}(\boldsymbol{W})$ by
\[
\widehat{\boldsymbol{A}}_{N}\left(\boldsymbol{\theta}^{\dagger},\boldsymbol{W}\right)=\left\{ \widehat{\boldsymbol{G}}_{N}\left(\boldsymbol{\theta}^{\dagger}\right)^{\top}\boldsymbol{W}\widehat{\boldsymbol{G}}_{N}\left(\boldsymbol{\theta}^{\dagger}\right)\right\} ^{-1}\widehat{\boldsymbol{G}}_{N}\left(\boldsymbol{\theta}^{\dagger}\right)^{\top}\boldsymbol{W}.
\]
The sandwich estimator is 
\begin{equation}
\widehat{\boldsymbol{\Sigma}}_{\mathrm{adj}}\left(\boldsymbol{\theta}^{\dagger},\boldsymbol{W}\right)=\frac{1}{N}\widehat{\boldsymbol{A}}_{N}\left(\boldsymbol{\theta}^{\dagger},\boldsymbol{W}\right)\widehat{\boldsymbol{C}}_{N}\left(\boldsymbol{\theta}^{\dagger}\right)\widehat{\boldsymbol{A}}_{N}\left(\boldsymbol{\theta}^{\dagger},\boldsymbol{W}\right)^{\top}.\label{eq:sandwich-estimator}
\end{equation}
This estimator targets the repeated-sampling covariance of the quasi-posterior
mean $\boldsymbol{\mu}_{N}(\boldsymbol{W})$ to first order, while
leaving the simulated quasi-posterior and its center unchanged. It
therefore does not alter the quasi-posterior itself or the finite-sample
influence of the prior on its center. This motivates calibrating $\boldsymbol{W}$
before the final quasi-posterior is simulated.

\subsection{Curvature-calibrated updating}

\label{subsec:updating}

Throughout this subsection, $\boldsymbol{\theta}$ denotes either
the parameter coordinate of a quasi-posterior or an MCMC draw. The
symbol $\boldsymbol{\vartheta}$ denotes a candidate calibration center.
Exact data-dependent quantities carry the subscript $N$, and a tilde
denotes a numerical approximation computed by MCMC.

For a candidate calibration center $\boldsymbol{\vartheta}\in\Theta$,
define 
\[
\boldsymbol{W}_{N}(\boldsymbol{\vartheta})=\widehat{\boldsymbol{C}}_{N}(\boldsymbol{\vartheta})^{-1}.
\]
Recall that $\boldsymbol{\mu}_{N}(\boldsymbol{W})$ denotes the exact
mean of the fixed-weight quasi-posterior conditional on the observed
data. Define the ideal calibration map by 
\[
\mathcal{T}_{N}(\boldsymbol{\vartheta})=\boldsymbol{\mu}_{N}\{\boldsymbol{W}_{N}(\boldsymbol{\vartheta})\}=\mathbb{E}_{\Pi^{\boldsymbol{W}_{N}(\boldsymbol{\vartheta})}(\cdot\mid\mathcal{D}_{N})}(\boldsymbol{\theta}).
\]
The ideal fixed-point formulation does not itself require a separate
consistent preliminary estimator. Let $\boldsymbol{\vartheta}^{(0)}_{N}$
be an initial center in the local contraction region characterized
below, and define the ideal calibration iterates recursively by 
\[
\boldsymbol{\vartheta}^{(s+1)}_{N}=\mathcal{T}_{N}\left(\boldsymbol{\vartheta}^{(s)}_{N}\right),\qquad s=0,1,\dots.
\]
We define the curvature-calibrated center as the local fixed point
$\boldsymbol{\vartheta}^{\star}_{N}=\mathcal{T}_{N}\left(\boldsymbol{\vartheta}^{\star}_{N}\right)$.
This condition is self-consistent: the covariance of the moments is
evaluated at the center of the quasi-posterior generated by the resulting
weighting matrix.

A convenient way to obtain an initial center without preliminary optimization
is through a pilot fixed-weight quasi-posterior. Choose a possibly
data-dependent positive-definite pilot weighting matrix $\boldsymbol{W}^{\mathrm{p}}_{N}$
that does not depend on the candidate calibration center. A deterministic
choice is included as a special case; for example, $\boldsymbol{W}^{\mathrm{p}}_{N}=\boldsymbol{I}_{K}$
may be used after rescaling moments whose numerical magnitudes differ
substantially. Let $\boldsymbol{\theta}^{\mathrm{p}}_{1},\dots,\boldsymbol{\theta}^{\mathrm{p}}_{T_{\mathrm{p}}}$
denote retained draws from $\Pi^{\boldsymbol{W}^{\mathrm{p}}_{N}}(d\boldsymbol{\theta}\mid\mathcal{D}_{N})$
and set 
\begin{equation}
\widetilde{\boldsymbol{\vartheta}}^{(0)}_{N}=\widetilde{\boldsymbol{\mu}}^{\mathrm{p}}_{N,T_{\mathrm{p}}}=\frac{1}{T_{\mathrm{p}}}\sum^{T_{\mathrm{p}}}_{t=1}\boldsymbol{\theta}^{\mathrm{p}}_{t}.\label{eq:pilot-initial-center}
\end{equation}
The pilot matrix need not be an estimate of $\boldsymbol{C}^{-1}_{0}$.
It suffices for the corresponding exact pilot quasi-posterior mean
to concentrate at $\boldsymbol{\theta}_{0}$ and for the pilot Monte
Carlo error to be $o_{p}(1)$. The pilot Markov chain may be initialized
at any state in a class over which this Monte Carlo approximation
holds uniformly. In particular, its initial state is unrestricted
when the approximation is uniform over the whole state space. Formal
conditions are stated in the Appendix.

The map $\mathcal{T}_{N}$ is generally unavailable in closed form.
For a candidate center $\boldsymbol{\vartheta}$ at calibration run
$s$, let $\boldsymbol{\theta}^{(s)}_{1},\dots,\boldsymbol{\theta}^{(s)}_{T_{s}}$
denote retained draws from $\Pi^{\boldsymbol{W}_{N}(\boldsymbol{\vartheta})}(d\boldsymbol{\theta}\mid\mathcal{D}_{N})$.
Define the Monte Carlo approximation to the map by 
\[
\widetilde{\mathcal{T}}^{(s)}_{N,T_{s}}(\boldsymbol{\vartheta})=\frac{1}{T_{s}}\sum^{T_{s}}_{t=1}\boldsymbol{\theta}^{(s)}_{t}.
\]
The computed iterates satisfy 
\[
\widetilde{\boldsymbol{\vartheta}}^{(s+1)}_{N}=\widetilde{\mathcal{T}}^{(s)}_{N,T_{s}}\left(\widetilde{\boldsymbol{\vartheta}}^{(s)}_{N}\right).
\]

We use two scale-free diagnostics to distinguish numerical error in
evaluating the calibration map from proximity to its fixed point.
First, define the reference covariance matrix 
\[
\widehat{\boldsymbol{\Sigma}}^{\mathrm{ref}}_{N}=\frac{1}{N}\left\{ \widehat{\boldsymbol{G}}_{N}\left(\widetilde{\boldsymbol{\vartheta}}^{(0)}_{N}\right)^{\top}\widehat{\boldsymbol{C}}_{N}\left(\widetilde{\boldsymbol{\vartheta}}^{(0)}_{N}\right)^{-1}\widehat{\boldsymbol{G}}_{N}\left(\widetilde{\boldsymbol{\vartheta}}^{(0)}_{N}\right)\right\} ^{-1}.
\]
For two candidate centers, define 
\[
\eta(\boldsymbol{u},\boldsymbol{v})=\left[\frac{1}{J}(\boldsymbol{u}-\boldsymbol{v})^{\top}\left(\widehat{\boldsymbol{\Sigma}}^{\mathrm{ref}}_{N}\right)^{-1}(\boldsymbol{u}-\boldsymbol{v})\right]^{1/2}.
\]
When the pilot center is consistent, this quantity measures the root-mean-square
change in the calibration center in asymptotic standard-error units.
More generally, the stopping argument only requires the eigenvalues
of $N\widehat{\boldsymbol{\Sigma}}^{\mathrm{ref}}_{N}$ to be bounded
above and bounded away from zero in probability. The criterion is
invariant under nonsingular linear reparameterizations.

Second, we assess the Monte Carlo accuracy with which each application
of the calibration map is evaluated. Under an MCMC central limit theorem,
the asymptotic covariance matrix of the chain mean can be estimated
using multivariate batch means \citep{Flegal2010,Flegal2015,Vats2019}.
For run $s$, we use the batch length 
\[
b_{s}=\left\lfloor \sqrt{T_{s}}\right\rfloor 
\]
and let $\widehat{\boldsymbol{\Omega}}^{(s)}_{\mathrm{BM}}$ denote
the corresponding estimator of the asymptotic covariance matrix in
the Markov-chain central limit theorem. The estimated Monte Carlo
covariance matrix of the chain mean is then 
\[
\widehat{\boldsymbol{V}}^{(s)}_{\mathrm{BM}}=\frac{1}{T_{s}}\widehat{\boldsymbol{\Omega}}^{(s)}_{\mathrm{BM}}.
\]
In particular, the componentwise Monte Carlo standard errors are 
\[
\widehat{\operatorname{MCSE}}^{(s)}_{j}=\left\{ \left(\widehat{\boldsymbol{V}}^{(s)}_{\mathrm{BM}}\right)_{jj}\right\} ^{1/2},\qquad j=1,\ldots,J.
\]
To summarize the Monte Carlo error of the entire mean vector on the
same scale as the fixed-point diagnostic, define 
\[
\chi^{(s)}=\left[\frac{1}{J}\operatorname{tr}\left\{ \left(\widehat{\boldsymbol{\Sigma}}^{\mathrm{ref}}_{N}\right)^{-1}\widehat{\boldsymbol{V}}^{(s)}_{\mathrm{BM}}\right\} \right]^{1/2}.
\]
Thus, $\chi^{(s)}$ is the root-mean-square Monte Carlo standard error
of the estimated quasi-posterior mean in reference standard-error
units. Like $\eta$, it is invariant under nonsingular linear reparameterizations.
Controlling $\chi^{(s)}$ limits the influence of Monte Carlo variability
on comparisons of consecutive calibration centers. The Monte Carlo
accuracy diagnostic relies on the validity of the Markov chain central
limit theorem and the adequacy of the batch-means variance estimator.
Satisfying the proposed stopping criteria therefore does not replace
ordinary MCMC convergence diagnostics.

The fixed point is approximated using the following steps.
\begin{enumerate}
\item Choose a positive-definite pilot weighting matrix $\boldsymbol{W}^{\mathrm{p}}_{N}$
that does not depend on the candidate calibration center, run the
pilot Markov chain, and set 
\[
\widetilde{\boldsymbol{\vartheta}}^{(0)}_{N}=\widetilde{\boldsymbol{\mu}}^{\mathrm{p}}_{N,T_{\mathrm{p}}}=\frac{1}{T_{\mathrm{p}}}\sum^{T_{\mathrm{p}}}_{t=1}\boldsymbol{\theta}^{\mathrm{p}}_{t}.
\]
Alternatively, any initial center that lies in the local contraction
region with probability approaching one may be used. If the pilot
center is to enter the first stopping comparison, extend the pilot
run as necessary so that its standardized Monte Carlo error satisfies
the same accuracy requirement described below.
\item At calibration iteration $s$, set 
\[
\boldsymbol{W}^{(s)}_{N}=\widehat{\boldsymbol{C}}_{N}\left(\widetilde{\boldsymbol{\vartheta}}^{(s)}_{N}\right)^{-1},
\]
run the corresponding fixed-weight Markov chain, and compute 
\[
\widetilde{\boldsymbol{\vartheta}}^{(s+1)}_{N}=\frac{1}{T_{s}}\sum^{T_{s}}_{t=1}\boldsymbol{\theta}^{(s)}_{t}
\]
together with $\chi^{(s)}$.
\item If $\chi^{(s)}>\tau_{\chi}$, extend the current chain and recompute
both $\widetilde{\boldsymbol{\vartheta}}^{(s+1)}_{N}$ and $\chi^{(s)}$.
Under the working approximation that the Monte Carlo standard error
decreases at rate $T^{-1/2}_{s}$, a convenient extrapolation for
the required total number of retained draws is 
\[
T^{\mathrm{new}}_{s}=\max\left\{ T_{s},\,\left\lceil T_{s}\left(\frac{\chi^{(s)}}{\tau_{\chi}}\right)^{2}\right\rceil \right\} .
\]
Because the asymptotic-variance estimate itself changes as the chain
is extended, this expression is used only to select the next chain
length; the MCSE criterion is subsequently re-evaluated. Repeat this
step until $\chi^{(s)}\leq\tau_{\chi}$.
\item Once the Monte Carlo accuracy criterion is satisfied, stop the calibration
iteration if 
\[
\eta\left(\widetilde{\boldsymbol{\vartheta}}^{(s+1)}_{N},\widetilde{\boldsymbol{\vartheta}}^{(s)}_{N}\right)\leq\tau_{\eta}.
\]
Otherwise, increment $s$ and return to Step 2.
\item Set $S=s+1$, define 
\[
\boldsymbol{W}^{(S)}_{N}=\widehat{\boldsymbol{C}}_{N}\left(\widetilde{\boldsymbol{\vartheta}}^{(S)}_{N}\right)^{-1},
\]
and run one final Markov chain targeting $\Pi^{\boldsymbol{W}^{(S)}_{N}}(d\boldsymbol{\theta}\mid\mathcal{D}_{N})$.
Retain the draws from this final run and assess their Monte Carlo
precision separately for the posterior summaries reported in the analysis. 
\end{enumerate}
Thus, $\tau_{\chi}$ controls the numerical precision with which each
application of the calibration map is evaluated, whereas $\tau_{\eta}$
controls proximity of consecutive calibration centers. For finite-sample
computation, $\tau_{\chi}$ should be chosen appreciably smaller than
$\tau_{\eta}$, so that Monte Carlo variability has little influence
on whether the fixed-point criterion is satisfied.

A fixed positive value of $\tau_{\chi}$ should be interpreted as
a finite-sample numerical tolerance. In particular, because $\widehat{\boldsymbol{\Sigma}}^{\mathrm{ref}}_{N}$
is of order $N^{-1}$, a fixed standardized tolerance controls the
Monte Carlo error to be a small constant fraction of an $N^{-1/2}$
sampling standard error but does not by itself imply $o_{p}\left(N^{-1/2}\right)$
Monte Carlo error. Whenever first-order equivalence of a computed
calibration iterate and its ideal counterpart is invoked asymptotically,
the simulation effort is therefore understood to increase so that
$\chi^{(s)}=o_{p}(1)$; equivalently, this may be formalized using
a sequence of computational tolerances $\tau_{\chi,N}\to0$.

We choose the Monte Carlo tolerance relative to the fixed-point tolerance.
Specifically, let $0<c_{\tau}<1$ denote the fraction of the fixed-point
stopping radius allocated to Monte Carlo uncertainty. If consecutive
calibration runs have standardized root-mean-square Monte Carlo errors
bounded by $\tau_{\chi}$, the Monte Carlo contribution to their difference
is approximately bounded in root-mean-square terms by $\sqrt{2}\tau_{\chi}$.
We therefore set 
\[
\tau_{\chi}=\frac{c_{\tau}}{\sqrt{2}}\tau_{\eta}.
\]
In our implementation we set $c_{\tau}=0.20$, allocating approximately
one fifth of the fixed-point stopping tolerance to Monte Carlo error
in root-mean-square terms under the working cross-run approximation.
This allocation is a numerical heuristic rather than a probabilistic
error bound. For $\tau_{\eta}=0.05$, this gives $\tau_{\chi}\simeq0.0071$.
The constant $c_{\tau}$ specifies a numerical error budget rather
than a statistical tuning parameter; smaller values increase computational
cost without altering the target distribution. 

The fixed tolerance above is used for finite-sample computation. For
asymptotic statements concerning equivalence between the computed
and ideal calibration iterates, simulation effort is understood to
increase with $N$ so that $\tau_{\chi,N}\to0$, and hence the Monte
Carlo error is negligible relative to the $N^{-1/2}$ statistical
scale.

Because the resulting weight calibrates the local quadratic curvature
of the criterion to the optimal GMM information matrix, we call the
procedure curvature-calibrated quasi-Bayesian (CCQB) updating. Since
the weighting matrix is fixed within each MCMC run, the procedure
avoids repeated evaluation of $\widehat{\boldsymbol{C}}_{N}(\boldsymbol{\theta})$
and its determinant. The iteration is used only to determine the locally
calibrated weighting matrix.

Two distinctions between the calibrated procedure and the fully parameter-dependent
quasi-posterior in \eqref{eq:quasi-posterior} are important. That
target contains the determinant term $\det\{\boldsymbol{W}_{N}(\boldsymbol{\theta})\}^{1/2}$
and allows the weighting matrix to vary with $\boldsymbol{\theta}$,
whereas \eqref{eq:fixed-W-quasi-posterior} fixes the weight and therefore
has a constant determinant. The calibrated fixed-weight quasi-posterior
matches the leading local quadratic term obtained by evaluating the
covariance weight at a self-consistent center, but it does not reproduce
variation in $\boldsymbol{W}_{N}(\boldsymbol{\theta})$ away from
that center or the nonconstant log-determinant contribution. Those
differences can affect higher-order and finite-sample features. The
computational advantage of fixing the weight within each MCMC run
must be balanced against the cost of additional calibration runs.
Since we do not provide a direct comparison of overall computing times,
we do not claim that CCQB is uniformly more computationally efficient
than alternative covariance-weighting procedures.

\subsection{Large-sample properties}

\label{subsec:theory}

The theorem below establishes three points needed for the proposed
workflow. First, the ideal calibration map is locally contractive
and therefore has a unique local fixed point. Second, only one or
two ideal updates are needed for first-order equivalence to that fixed
point, depending on the quality of the initial center. Third, a first-order
mean expansion and Bernstein--von Mises approximation established
at the efficient population weight $\boldsymbol{W}_{0}$ continue
to hold under the self-consistently calibrated random weight.

The identification, smoothness, and positive-definiteness conditions
are local to a neighborhood of the true parameter. The result also
imposes uniform moment bounds on the fixed-weight quasi-posteriors
and assumes a Bernstein--von Mises approximation at the population-optimal
weight. These additional conditions concern posterior concentration
and tail behavior, rather than only local properties of the moment
functions. Conditional on these properties, the theorem establishes
the stability and calibration of the self-consistent procedure. Its
requirements include local identification; smoothness and positive
definiteness of $\boldsymbol{C}(\boldsymbol{\theta})$; regularity
of the estimated weighting map and fixed-weight quasi-posteriors;
and a prior that is positive near $\boldsymbol{\theta}_{0}$. It does
not require global identification or consistency of an arbitrary initial
center: the ideal iteration only needs to start in the local contraction
region with probability approaching one. A separate global concentration
condition justifies the pilot construction. Formal conditions and
the high-probability contraction event are stated in the Appendix.
Stochastic orders for exact data-dependent quantities refer to the
sampling law; statements involving MCMC output refer to the joint
law of the data and simulation.

\begin{theorem}

\label{thm:main-thm}

Suppose that Assumption~\ref{ass:regularity-cond} in the Appendix
holds. There exist a fixed $\delta>0$ and a sequence of events $\mathcal{E}_{N}$
satisfying $\Pr(\mathcal{E}_{N})\to1$ such that, on $\mathcal{E}_{N}$,
the calibration map 
\[
\mathcal{T}_{N}(\boldsymbol{\vartheta})=\boldsymbol{\mu}_{N}\{\boldsymbol{W}_{N}(\boldsymbol{\vartheta})\}
\]
maps 
\[
\overline{\mathcal{B}}_{\delta}(\boldsymbol{\theta}_{0})=\left\{ \boldsymbol{\vartheta}\in\mathbb{R}^{J}:\lVert\boldsymbol{\vartheta}-\boldsymbol{\theta}_{0}\rVert\leq\delta\right\} 
\]
into itself and is a contraction on this set. Its contraction modulus
may be chosen to satisfy 
\[
\rho_{N}=O_{p}\left(N^{-1/2}\right).
\]
On $\mathcal{E}_{N}$, let $\boldsymbol{\vartheta}^{\star}_{N}$ denote
the unique fixed point. Extend it by setting 
\[
\boldsymbol{\vartheta}^{\star}_{N}=\boldsymbol{\mu}_{N}(\boldsymbol{W}_{0})\quad\text{on }\mathcal{E}^{c}_{N},
\]
and define 
\[
\boldsymbol{W}^{\star}_{N}=\begin{cases}
\boldsymbol{W}_{N}\left(\boldsymbol{\vartheta}^{\star}_{N}\right), & \text{on }\mathcal{E}_{N},\\
\boldsymbol{W}_{0}, & \text{on }\mathcal{E}^{c}_{N}.
\end{cases}
\]
The ideal iteration converges to $\boldsymbol{\vartheta}^{\star}_{N}$
from every starting value in the ball on $\mathcal{E}_{N}$. Hence,
for any possibly random starting sequence satisfying 
\[
\Pr\left\{ \boldsymbol{\vartheta}^{(0)}_{N}\in\overline{\mathcal{B}}_{\delta}(\boldsymbol{\theta}_{0})\right\} \to1,
\]
the iteration converges to $\boldsymbol{\vartheta}^{\star}_{N}$ with
probability approaching one. Moreover, 
\[
\boldsymbol{\vartheta}^{(1)}_{N}-\boldsymbol{\theta}_{0}=O_{p}\left(N^{-1/2}\right),\qquad\boldsymbol{\vartheta}^{(1)}_{N}-\boldsymbol{\vartheta}^{\star}_{N}=O_{p}\left(N^{-1/2}\right),
\]
and 
\[
\boldsymbol{\vartheta}^{(2)}_{N}-\boldsymbol{\vartheta}^{\star}_{N}=O_{p}\left(N^{-1}\right)=o_{p}\left(N^{-1/2}\right).
\]
If, in addition, 
\[
\boldsymbol{\vartheta}^{(0)}_{N}\xrightarrow{p}\boldsymbol{\theta}_{0},
\]
then 
\[
\boldsymbol{\vartheta}^{(1)}_{N}-\boldsymbol{\vartheta}^{\star}_{N}=o_{p}\left(N^{-1/2}\right).
\]
The fixed point and calibrated weight satisfy 
\[
\boldsymbol{\vartheta}^{\star}_{N}-\boldsymbol{\theta}_{0}=O_{p}\left(N^{-1/2}\right),\qquad\boldsymbol{W}^{\star}_{N}\xrightarrow{p}\boldsymbol{W}_{0},
\]
and 
\[
\sqrt{N}\left(\boldsymbol{\vartheta}^{\star}_{N}-\boldsymbol{\theta}_{0}\right)=-\boldsymbol{A}_{0}\boldsymbol{\xi}_{N}+o_{p}(1).
\]
Conditionally on the data, 
\[
d_{\mathrm{BL}}\left[\mathcal{L}_{\Pi^{\boldsymbol{W}^{\star}_{N}}(\cdot\mid\mathcal{D}_{N})}\left\{ \sqrt{N}\left(\boldsymbol{\theta}-\boldsymbol{\vartheta}^{\star}_{N}\right)\right\} ,\;\mathcal{N}(\boldsymbol{0}_{J},\boldsymbol{V}_{0})\right]\xrightarrow{p}0,
\]
where $d_{\mathrm{BL}}(\cdot,\cdot)$ denotes bounded-Lipschitz distance.
Moreover, 
\[
N\operatorname{Var}_{\Pi^{\boldsymbol{W}^{\star}_{N}}(\cdot\mid\mathcal{D}_{N})}(\boldsymbol{\theta})\xrightarrow{p}\boldsymbol{V}_{0}.
\]
If, in addition, 
\[
\sqrt{N}\bar{\boldsymbol{m}}_{N}(\boldsymbol{\theta}_{0})\rightsquigarrow\mathcal{N}(\boldsymbol{0}_{K},\boldsymbol{C}_{0}),
\]
then 
\[
\sqrt{N}\left(\boldsymbol{\vartheta}^{\star}_{N}-\boldsymbol{\theta}_{0}\right)\rightsquigarrow\mathcal{N}(\boldsymbol{0}_{J},\boldsymbol{V}_{0}).
\]

\end{theorem}

The proof is given in the Appendix.

The conditional Gaussian approximation and the limiting distribution
of $\boldsymbol{\vartheta}^{\star}_{N}$ have the same first-order
covariance matrix $\boldsymbol{V}_{0}$. Thus, at the level of the
first-order Gaussian approximation, the calibrated quasi-posterior
dispersion and the asymptotic sampling dispersion of its center are
governed by the inverse Godambe information matrix associated with
the specified moments.

The contraction modulus is $O_{p}\left(N^{-1/2}\right)$. Therefore,
the first ideal update maps any initial center in the local contraction
region into an $O_{p}\left(N^{-1/2}\right)$ neighborhood of $\boldsymbol{\theta}_{0}$,
and the second update is first-order equivalent to the fixed point.
When the initial center is consistent---as is the pilot MCMC center
under the conditions in the Appendix---one ideal calibration update
is already first-order equivalent. The pilot run itself only needs
$o_{p}(1)$ Monte Carlo accuracy, whereas the calibration runs require
Monte Carlo error to be negligible relative to $N^{-1/2}$ for first-order
equivalence of the computed iterate.

The proposed procedure should not be viewed as a Bayesian reinterpretation
of iterated GMM \citep{Hansen1996}. The overall scale of $\boldsymbol{W}$
acts as a learning rate, whereas its normalized shape determines the
relative weighting of the moment conditions. The prior may still have
an important finite-sample effect, but the empirical component is
no longer placed on an arbitrary local scale.

\section{Simulation study}

\label{sec:simulation}

\subsection{Questions addressed by the simulation}

\label{subsec:sim-purpose}

The simulation study examines three related questions. First, we assess
whether replacing an identity weighting matrix with a covariance-based
weight improves the agreement between quasi-posterior uncertainty
and the repeated-sampling variability of its center. Second, conditional
on covariance weighting, we investigate whether iteration to the self-consistent
CCQB fixed point provides additional finite-sample improvements over
a single covariance update. Third, we compare CCQB with a sample-oracle
covariance-weighted procedure, which evaluates the estimated moment
covariance at the true parameter, and Bayesian exponentially tilted
empirical likelihood \citep[BETEL;][]{Chib2018}, a leading alternative
for Bayesian inference under moment restrictions.

The primary evaluation criteria are the coverage and mean lengths
of nominal $95\%$ intervals, together with discrepancies between
reported and repeated-sampling covariance matrices. These measures
focus on uncertainty calibration rather than ranking procedures by
the bias or mean squared error of their centers.

\subsection{Designs and common settings}

\label{subsec:mc-design}

We consider three data-generating processes: an overidentified instrumental-variables
(IV) regression design, a heteroskedastic and serially dependent (HSD)
linear regression design, and a Poisson regression design with a nonlinear
moment function. In every design, the parameter dimension is $J=4$,
the sample size is
\[
N\in\{50,100,200,500\},
\]
and the prior is
\[
\boldsymbol{\theta}\sim\mathcal{N}\left(\boldsymbol{0}_{J},\varrho^{2}\boldsymbol{I}_{J}\right),\qquad\varrho=5.
\]
Unless otherwise stated explicitly, primitive covariates and disturbances
are independent across observations and mutually independent.

The moment vector $\boldsymbol{m}_{i}(\boldsymbol{\theta})$ is specified
separately for each design. In the IV design, let $\tilde{x}_{i}$
denote the endogenous regressor, $\boldsymbol{w}_{i}\in\mathbb{R}^{3}$
the included exogenous regressors, and $\boldsymbol{q}_{i}\in\mathbb{R}^{5}$
the excluded instruments. With $\boldsymbol{x}_{i}=(\tilde{x}_{i},\boldsymbol{w}^{\top}_{i})^{\top}$
and $\boldsymbol{z}_{i}=(\boldsymbol{q}^{\top}_{i},\boldsymbol{w}^{\top}_{i})^{\top}$,
the outcome and moment vector are
\[
y_{i}=\boldsymbol{x}^{\top}_{i}\boldsymbol{\theta}_{0}+\varepsilon_{i},\qquad\boldsymbol{m}_{i}(\boldsymbol{\theta})=(y_{i}-\boldsymbol{x}^{\top}_{i}\boldsymbol{\theta})\boldsymbol{z}_{i}.
\]
The first-stage disturbance in $\tilde{x}_{i}$ is correlated with
the outcome disturbance $\varepsilon_{i}$. In the Poisson design,
\[
y_{i}\mid\boldsymbol{x}_{i}\sim\operatorname{Poisson}\left\{ \exp(\boldsymbol{x}^{\top}_{i}\boldsymbol{\theta}_{0})\right\} ,\qquad\boldsymbol{m}_{i}(\boldsymbol{\theta})=\left\{ y_{i}-\exp(\boldsymbol{x}^{\top}_{i}\boldsymbol{\theta})\right\} \boldsymbol{x}_{i}.
\]
The covariates in this design are Gaussian. Finally, the HSD design
uses $y_{i}=\boldsymbol{x}^{\top}_{i}\boldsymbol{\theta}_{0}+\sigma_{i}u_{i}$
and $\boldsymbol{m}_{i}(\boldsymbol{\theta})=(y_{i}-\boldsymbol{x}^{\top}_{i}\boldsymbol{\theta})\boldsymbol{x}_{i}$,
with covariate-dependent $\sigma^{2}_{i}$ and serial dependence in
both the regressors and $u_{i}$. Its covariance weight is estimated
using a Newey--West long-run covariance estimator.

The HSD and Poisson designs are exactly identified. When $\boldsymbol{G}_{0}$
is nonsingular,
\[
\boldsymbol{A}(\boldsymbol{W})=\left(\boldsymbol{G}^{\top}_{0}\boldsymbol{W}\boldsymbol{G}_{0}\right)^{-1}\boldsymbol{G}^{\top}_{0}\boldsymbol{W}=\boldsymbol{G}^{-1}_{0}.
\]
Consequently, the weighting matrix does not affect the first-order
sampling distribution of the quasi-posterior center. It does, however,
affect the covariance represented by the quasi-posterior and may affect
its center through higher-order effects and its finite-sample interaction
with the prior. The exactly identified designs therefore isolate covariance
calibration and higher-order effects from first-order changes in the
sampling distribution of the center. By contrast, in the overidentified
IV design the relative weighting of the moment conditions can also
affect the first-order sampling distribution of the center.

The main text focuses on two complementary designs: the overidentified
IV design with $K=8$ and the exactly identified Poisson design with
$K=4$. The IV design has a quadratic criterion and a Gaussian fixed-weight
quasi-posterior available in closed form. It therefore provides a
transparent benchmark for studying the effect of changing the weighting
matrix without numerical simulation error. By contrast, the Poisson
design combines exact identification with a nonlinear moment function
and a non-Gaussian quasi-posterior. It therefore provides a setting
in which higher-order finite-sample effects of self-consistent calibration
may be more pronounced. Section~S1 of the Supplementary Material
gives the design and computational details, and Section~S2 reports
the additional results.

\subsection{Competing procedures}

\label{subsec:compet-proc}

We compare three stages of CCQB calibration, indexed by $s\in\{0,1,\star\}$,
and two benchmarks. In the evaluation below, $a\in\{0,1,\star,\mathrm{or},\mathrm{BETEL}\}$
indexes all five procedures. Raw and sandwich-adjusted summaries are
available for the quadratic quasi-posteriors, while BETEL has only
posterior summaries. The comparison between $s=0$ and $s=1$ isolates
the effect of replacing the identity weight by a covariance weight,
whereas the comparison between $s=1$ and $s=\star$ isolates the
additional finite-sample effect of imposing self-consistency.

For each replication, we first evaluate the identity-weight pilot
quasi-posterior, denoted by $s=0$, with $\boldsymbol{W}^{p}=\boldsymbol{I}_{K}$.
The one-step covariance-weighted procedure, denoted by $s=1$, then
fixes $\boldsymbol{W}^{(0)}_{N}=\widehat{\boldsymbol{C}}_{N}(\widetilde{\boldsymbol{\vartheta}}^{(0)}_{N})^{-1}$
using the center of the pilot quasi-posterior. Finally, $s=\star$
denotes the converged CCQB procedure, for which the covariance weight
is repeatedly updated until the stopping criterion is satisfied. For
each weighting scheme, we report both the raw quasi-posterior uncertainty
and its sandwich-adjusted counterpart. These are alternative uncertainty
summaries for the same quasi-posterior center rather than separate
point-estimation procedures. In particular, the sandwich adjustment
changes the reported covariance but leaves both the underlying quasi-posterior
and its center unchanged. 

As a benchmark for covariance weighting, we consider the sample-oracle
weight $\boldsymbol{W}^{\mathrm{or}}_{N}=\widehat{\boldsymbol{C}}_{N}(\boldsymbol{\theta}_{0})^{-1}$.
It uses the true parameter as the covariance-evaluation point but
retains the sampling variability of $\widehat{\boldsymbol{C}}_{N}$,
and should therefore be distinguished from the population-optimal
weight $\boldsymbol{C}^{-1}_{0}$. We also include BETEL as a nonquadratic
moment-based Bayesian comparator, using the same moment restrictions
and parameter prior \citep{Chib2018}. The sample-oracle is summarized
by both raw and sandwich-adjusted uncertainty; BETEL is summarized
by its posterior quantile intervals and posterior covariance. We omit
observation-level BETEL from the serially dependent HSD design because
that implementation does not account for the long-run moment covariance.
Computational details, including the treatment of dependence for the
HSD design, are given in Section~S1 of the Supplementary Material.

Each design uses $R=1{,}000$ Monte Carlo replications. CCQB uses
the standardized fixed-point tolerance $\tau_{\eta}=0.05$ and the
Monte Carlo tolerance $\tau_{\chi}=(0.20/\sqrt{2})\tau_{\eta}\simeq0.0071$.
The latter requirement is automatically satisfied when the calibration
map is evaluated analytically. For each method, we record failures
arising from non-positive-definite covariance estimates, numerical
errors, or failure to satisfy the CCQB stopping rule within 100 updates.
BETEL was successfully evaluated in all replications across all sample
sizes in the IV and Poisson designs. Comparisons across methods are
restricted to replications in which all methods involved in the comparison
are successfully evaluated.

\subsection{Evaluation measures}

\label{subsec:eval-crit}

Let $\widehat{\boldsymbol{\mu}}_{a,r}$ denote the quasi-posterior
mean and let $\widehat{\boldsymbol{\Sigma}}^{\mathrm{raw}}_{a,r}$
denote its covariance estimate. For the quadratic quasi-posteriors,
$\widehat{\boldsymbol{\Sigma}}^{\mathrm{adj}}_{a,r}$ denotes the
sandwich covariance estimate at the same center and weighting matrix.
Hereafter, ``raw'' and ``adj'' refer to the unadjusted and sandwich-adjusted
uncertainty summaries, respectively. Let $q_{a,r,j}(p)$ denote the
marginal quasi-posterior quantile at probability $p$, computed analytically
for the Gaussian IV and HSD designs and empirically from the retained
MCMC draws for the Poisson design. The raw and sandwich-adjusted $95\%$
intervals are
\[
I^{\mathrm{raw}}_{a,r,j}=[q_{a,r,j}(0.025),q_{a,r,j}(0.975)]
\]
and
\[
I^{\mathrm{adj}}_{a,r,j}=\widehat{\mu}_{a,r,j}\pm\zeta_{0.975}\sqrt{\left(\widehat{\boldsymbol{\Sigma}}^{\mathrm{adj}}_{a,r}\right)_{jj}},
\]
respectively, where $\zeta_{0.975}$ denotes the $0.975$ quantile
of the standard normal distribution. For BETEL, only the raw interval
is used. 

We report Monte Carlo coverage probabilities and mean interval lengths
over the common set of successful replications. We interpret interval
length alongside coverage, rather than as a measure of point-estimation
efficiency. In the IV and HSD designs, the fixed-weight quasi-posteriors
are Gaussian, so differences between the raw and adjusted intervals
primarily reflect differences in the covariance used to construct
them. In the Poisson design, however, the raw intervals are quasi-posterior
quantile intervals and need not be symmetric, whereas the adjusted
intervals are symmetric normal intervals centered at the same quasi-posterior
mean. The raw-versus-adjusted coverage comparison for the Poisson
design therefore reflects both covariance adjustment and finite-sample
quasi-posterior shape. We consequently use the covariance-discrepancy
measures below as a more direct assessment of covariance calibration.

To evaluate covariance calibration, we define the following diagnostics.
For each comparison, let $\mathcal{R}$ be its common set of successful
replications. The repeated-sampling covariance of the quasi-posterior
center is estimated by 
\[
\widehat{\boldsymbol{V}}^{\mathrm{RS}}_{a}=\frac{1}{|\mathcal{R}|-1}\sum_{r\in\mathcal{R}}\left(\widehat{\boldsymbol{\mu}}_{a,r}-\bar{\boldsymbol{\mu}}_{a}\right)\left(\widehat{\boldsymbol{\mu}}_{a,r}-\bar{\boldsymbol{\mu}}_{a}\right)^{\top},
\]
with 
\[
\bar{\boldsymbol{\mu}}_{a}=\frac{1}{|\mathcal{R}|}\sum_{r\in\mathcal{R}}\widehat{\boldsymbol{\mu}}_{a,r}.
\]
We use an affine-invariant covariance discrepancy. For $u\in\{\mathrm{raw},\mathrm{adj}\}$,
define 
\[
\boldsymbol{\Upsilon}^{u}_{a,r}=\left(\widehat{\boldsymbol{V}}^{\mathrm{RS}}_{a}\right)^{-1/2}\widehat{\boldsymbol{\Sigma}}^{u}_{a,r}\left(\widehat{\boldsymbol{V}}^{\mathrm{RS}}_{a}\right)^{-1/2},
\]
where $\boldsymbol{B}^{-1/2}$ denotes the symmetric inverse square
root of a matrix $\boldsymbol{B}$. Let $\lambda^{u}_{a,r,1},\dots,\lambda^{u}_{a,r,J}$
denote the eigenvalues of $\boldsymbol{\Upsilon}^{u}_{a,r}$. Equivalently,
these are the generalized eigenvalues of the covariance pair $\left(\widehat{\boldsymbol{\Sigma}}^{u}_{a,r},\widehat{\boldsymbol{V}}^{\mathrm{RS}}_{a}\right)$.
We measure covariance discrepancy by
\[
d^{u}_{\mathrm{cov},a,r}=\left\{ \frac{1}{J}\sum^{J}_{j=1}\left(\log\lambda^{u}_{a,r,j}\right)^{2}\right\} ^{1/2}
\]
and report its median
\[
D^{u}_{\mathrm{cov},a}=\operatorname{median}(d^{u}_{\mathrm{cov},a,r}).
\]
The generalized eigenvalues of $\left(\widehat{\boldsymbol{\Sigma}}^{u}_{a,r},\widehat{\boldsymbol{V}}^{\mathrm{RS}}_{a}\right)$
can be interpreted as directional variance ratios of the reported
covariance relative to the repeated-sampling covariance of the quasi-posterior
center. Thus, $D^{u}_{\mathrm{cov},a}$ provides an affine-invariant
summary of overall covariance miscalibration, where zero indicates
exact agreement.

\subsection{Simulation results}

\label{subsec:sim-results}

Table~\ref{tab:updates} shows rapid convergence under the stopping
rule. No CCQB calibration runs fail in either design. In the IV design,
the median number of updates is three at $N=50$ and two for $N\geq100$;
in the Poisson design, it is three throughout. As $N$ increases from
50 to 500, the maximum falls from eight to two in the IV design and
from seven to four in the Poisson design. Thus, self-consistent calibration
requires only a few updates in these designs.

\medskip{}

\begin{center}
{[}Table~\ref{tab:updates} here.{]}
\par\end{center}

\medskip{}

Table~\ref{tab:coverage} reports coverage for $\theta_{1}$. In
the IV design, the raw identity-weight intervals severely undercover,
whereas sandwich adjustment largely corrects the distortion. Covariance
weighting substantially improves raw coverage, with negligible differences
between $s=1$ and $s=\star$ or between raw and adjusted intervals.
Further iteration does not uniformly improve coverage across parameters.
In the Poisson design, for example, raw coverage for $\theta_{4}$
decreases from 0.878 under one-step weighting to 0.839 under converged
CCQB at $N=50$, and from 0.979 to 0.913 at $N=500$ (Table S3). Thus,
the improvements for $\theta_{1}$ should not be interpreted as evidence
that self-consistency always brings coverage closer to the nominal
level. The sample-oracle offers better raw coverage at small $N$,
while BETEL undercovers more markedly; the methods approach nominal
coverage as $N$ grows. In the Poisson design, raw coverage is 1.000
under $s=0$ and approximately 0.99 under $s=1$. Iteration reduces
it to 0.925--0.958 under $s=\star$, closer to the nominal 95\% level
and increasingly similar to the oracle. BETEL coverage rises from
0.900 to 0.949, but is lower in smaller samples.

\medskip{}

\begin{center}
{[}Table~\ref{tab:coverage} here.{]}
\par\end{center}

\medskip{}

Tables~S1 and S3 in Section~S2 of the Supplementary Material give
the parameter-wise coverage results. In the IV design, $s=1$ and
$s=\star$ agree closely across all coefficients; the oracle generally
improves small-sample raw coverage, while BETEL undercovers more strongly
at $N=50$. In the Poisson design, iteration reduces raw overcoverage
for the first three coefficients, although $\theta_{3}$ then undercovers
at smaller sample sizes. Calibration remains most difficult for $\theta_{4}$:
at $N=500$, converged raw and adjusted coverage are 0.913 and 0.921,
respectively.

Table~\ref{tab:cov-calib} assesses covariance calibration directly.
In the IV design, the raw discrepancy remains near three under $s=0$,
but falls substantially after one covariance update; further iteration
changes little. The sample-oracle has a slightly smaller discrepancy,
whereas BETEL is less well calibrated than the covariance-weighted
CCQB procedures. In the Poisson design, covariance weighting markedly
improves on $s=0$, and iteration further reduces the raw discrepancy
from 0.862 at $N=50$ to 0.306 at $N=500$. The latter is close to
the oracle (0.304) and BETEL (0.297); under $s=\star$, raw and adjusted
discrepancies at $N=500$ are also nearly identical (0.306 and 0.310).

\medskip{}

\begin{center}
{[}Table~\ref{tab:cov-calib} here.{]}
\par\end{center}

\medskip{}

The HSD results in Section~S2 of the Supplementary Material show
that the improvement in covariance calibration also extends to serially
dependent observations. At $N=500$, the median raw covariance discrepancy
falls from $3.687$ under $s=0$ to $0.583$ under $s=1$ and remains
$0.583$ under $s=\star$. For $\theta_{1}$, raw 95\% interval coverage
rises from $0.214$ under $s=0$ to $0.927$ under $s=1$ and is unchanged
under $s=\star$, although it remains below the nominal level. Thus,
accounting for the long-run moment covariance provides most of the
improvement in this design; further iteration has little additional
effect.

Table~\ref{tab:length} reports mean lengths of nominal 95\% intervals
for $\theta_{1}$. In the IV design, raw identity-weight intervals
are far too short, whereas $s=1$ and $s=\star$ have nearly identical
raw and adjusted lengths. Oracle intervals are somewhat longer at
small $N$; BETEL intervals are shorter, but exhibit greater undercoverage.
In the Poisson design, the very wide raw intervals under $s=0$ shorten
substantially after covariance weighting and further under $s=\star$.
Converged raw and adjusted lengths approach one another as $N$ grows
and are similar to those under the oracle. BETEL generally produces
the shortest intervals, with some loss of coverage. These length comparisons
should be read alongside Table~\ref{tab:coverage}.

\medskip{}

\begin{center}
{[}Table~\ref{tab:length} here.{]}
\par\end{center}

\medskip{}

\section{Real data application}

\label{sec:application}

\subsection{Data and empirical design}

\label{subsec:wheeze-setup}

We examine the age-specific association between baseline maternal
smoking status and childhood wheeze using the Ohio Children Wheeze
Status data from the Six Cities study \citep{Fitzmaurice1993}, available
in the R package \texttt{geepack}. The data contain four annual binary
wheeze measurements for each of $537$ children, observed from ages
$7$ to $10$. We treat children as the independent sampling units,
so that $N=537$. Of the $537$ children, $187$ had mothers classified
as smokers in the first study year (age $7$); the observed proportion
of children with wheeze was higher in this group at each of the four
ages.

Let $Y_{it}$ indicate whether child $i$ wheezes at visit $t$, let
$\texttt{age}_{it}$ denote age centered at $9$ years, and let $\texttt{smoke}_{i}$
indicate maternal smoking status at baseline. We posit the marginal
mean model 
\begin{align*}
p_{it}(\boldsymbol{\theta}) & =\Pr(Y_{it}=1\mid\texttt{age}_{it},\texttt{smoke}_{i})\\
 & =\operatorname{logit}^{-1}(\theta_{1}+\theta_{2}\texttt{age}_{it}+\theta_{3}\texttt{smoke}_{i}+\theta_{4}\texttt{age}_{it}\texttt{smoke}_{i}),
\end{align*}
where $\boldsymbol{\theta}=(\theta_{1},\theta_{2},\theta_{3},\theta_{4})^{\top}$.
Our primary estimand is the age-specific smoking-wheeze odds ratio
\[
\operatorname{OR}(a)=\exp(\theta_{3}+\theta_{4}(a-9)),\qquad a\in\{7,8,9,10\}.
\]
Thus, $\exp(\theta_{3})$ is the smoking odds ratio at age $9$, while
$\exp(\theta_{4})$ is the multiplicative change in this odds ratio
for each one-year increase in age.

We construct eight age-by-smoking residual moment conditions. Let
\[
\mathcal{A}=\{-2,-1,0,1\},
\]
corresponding to ages 7 through 10. For $g\in\{0,1\}$ and $r\in\mathcal{A}$,
define 
\[
m_{i,(g,r)}(\boldsymbol{\theta})=\mathbf{1}(\texttt{smoke}_{i}=g)\sum^{4}_{t=1}\mathbf{1}(\texttt{age}_{it}=r)\{Y_{it}-p_{it}(\boldsymbol{\theta})\}.
\]
Stacking these moments gives 
\[
\boldsymbol{m}_{i}(\boldsymbol{\theta})=\{m_{i,(g,r)}(\boldsymbol{\theta}):g\in\{0,1\},r\in\mathcal{A}\}^{\top}\in\mathbb{R}^{8}.
\]
Hence $J=4$ and $K=8$, yielding an overidentified moment model that
fits a four-parameter marginal trajectory to the eight age-by-smoking
cell means. For a candidate calibration center $\boldsymbol{\vartheta}$,
$\widehat{\boldsymbol{C}}_{N}(\boldsymbol{\vartheta})$ is the empirical
covariance of the child-level moment vectors, and 
\[
\boldsymbol{W}_{N}(\boldsymbol{\vartheta})=\widehat{\boldsymbol{C}}_{N}(\boldsymbol{\vartheta})^{-1}.
\]
Because the covariance is estimated across children using the full
child-level moment vector, within-child dependence among the four
repeated measurements is accommodated without specifying a working
correlation structure.

We assign informative priors to the smoking association and its age
interaction, while using relatively diffuse priors for the baseline
trajectory. Independently, we specify 
\[
\theta_{3}\sim\mathcal{N}\left(0,0.5^{2}\right),
\]
which assigns approximately 95\% prior probability to 
\[
0.38<\exp(\theta_{3})<2.66.
\]
For the interaction, we elicit the prior in terms of the change in
the smoking odds ratio over the observed age range: 
\[
\theta_{4}\sim\mathcal{N}\left(0,\left\{ \frac{\log2}{3(1.96)}\right\} ^{2}\right).
\]
Since 
\[
\frac{\operatorname{OR}(10)}{\operatorname{OR}(7)}=\exp(3\theta_{4}),
\]
this prior assigns approximately 95\% probability to 
\[
\frac{1}{2}<\frac{\operatorname{OR}(10)}{\operatorname{OR}(7)}<2.
\]
Thus, the prior favors gradual age variation, assigning approximately
$95\%$ probability to a change in the smoking odds ratio by no more
than a factor of two in either direction between ages $7$ and $10$.
Finally, we use weakly informative proper priors 
\[
\theta_{1},\theta_{2}\sim\mathcal{N}\left(0,5^{2}\right).
\]

We use the No-U-Turn sampler for each fixed-weight quasi-posterior,
following the computational approach described for the Poisson regression
design in Section~S1.3 of the Supplementary Material. The application-specific
computational settings are described below. We first run an identity-weight
pilot quasi-posterior with $\boldsymbol{W}^{\mathrm{p}}_{N}=\boldsymbol{I}_{8}$,
initializing the chain at a draw from the prior. Each calibration
chain is initialized at the estimated quasi-posterior mean of the
preceding chain, and its covariance weight is evaluated at that mean.
Each chain initially consists of $10{,}000$ warm-up iterations followed
by $20{,}000$ retained iterations. If the Monte Carlo accuracy requirement
is not met, the corresponding fixed-weight chain is rerun with a larger
number of retained iterations, keeping its weight, initial state,
random seed, and warm-up settings unchanged. Thus, $30{,}000$ iterations
is the initial run length, not a fixed length for every calibration
run.

We implement the two-part stopping rule in Section~\ref{subsec:updating}.
The reference covariance is computed at the pilot quasi-posterior
mean and kept fixed during calibration. We set the fixed-point tolerance
to $\tau_{\eta}=0.05$ and the standardized Monte Carlo accuracy tolerance
to $\tau_{\chi}=(0.20/\sqrt{2})\tau_{\eta}\simeq0.0071$. The pilot
and each calibration chain must satisfy the accuracy criterion $\chi\leq\tau_{\chi}$
before the displacement between successive calibration centers is
compared with $\tau_{\eta}$. Calibration terminates when this displacement
satisfies $\eta\leq\tau_{\eta}$. We then run a separate final chain
using the covariance weight evaluated at the terminal calibration
center; its Monte Carlo precision is assessed separately.

We report the identity-weight pilot ($s=0$), the first covariance-weight
update ($s=1$), and the final calibrated quasi-posterior ($s=\star$).
For each, we report quasi-posterior means and raw and sandwich-adjusted
$95\%$ intervals. Raw parameter intervals use the empirical $2.5\%$
and $97.5\%$ quantiles of the retained draws, while sandwich-adjusted
parameter intervals use the normal critical value $z_{0.975}=1.96$
and the estimated sandwich covariance. For the age-specific odds ratios,
we average the transformed draws for point estimates and use their
empirical $2.5\%$ and $97.5\%$ quantiles for raw intervals; the
adjusted intervals use a normal approximation with a delta-method
standard error on the odds-ratio scale. 

\subsection{Results}

\label{subsec:wheeze-results}

Table~\ref{tab:wheeze-theta} reports the parameter summaries under
the identity-weight pilot ($s=0$), the first covariance-weight update
($s=1$), and the final calibrated quasi-posterior ($s=\star$). Replacing
the pilot weight has a noticeable effect on both location and dispersion.
From $s=0$ to $s=1$, the quasi-posterior means of $\theta_{1}$,
$\theta_{3}$, and $\theta_{4}$ change from $-2.039$ to $-1.917$,
$0.081$ to $0.219$, and $0.001$ to $0.046$, respectively, whereas
the mean of $\theta_{2}$ changes only from $-0.141$ to $-0.135$.
Raw intervals also become substantially narrower for several parameters.
By comparison, the estimates at $s=1$ and $s=\star$ are very close:
the largest absolute change in a quasi-posterior mean is $0.003$
(for $\theta_{3}$), and the interval endpoints change little. Thus,
most of the change relative to the identity-weight pilot occurs upon
introducing covariance weighting, with little additional change before
the stopping criteria are met.

\medskip{}

\begin{center}
{[}Table~\ref{tab:wheeze-theta} here.{]}
\par\end{center}

\medskip{}

After calibration, the raw quasi-posterior quantile intervals and
the sandwich-adjusted normal intervals give broadly similar conclusions,
although the adjusted intervals are generally wider. The difference
is particularly visible for $\theta_{4}$: at $s=\star$, its raw
$95\%$ interval is $[-0.093,0.188]$, compared with $[-0.133,0.228]$
after sandwich adjustment. Both intervals for $\theta_{2}$ lie below
zero, whereas those for $\theta_{3}$ and $\theta_{4}$ include zero.
The remaining discrepancies are consistent with the first-order nature
of covariance calibration and with the different constructions of
quasi-posterior quantile and sandwich-adjusted normal intervals.

Table~\ref{tab:wheeze-or} reports the corresponding age-specific
smoking--wheeze odds ratios. Under the final calibrated quasi-posterior,
their means are $1.158$, $1.210$, $1.270$, and $1.340$ at ages
$7$, $8$, $9$, and $10$, respectively. The identity-weight pilot
instead gives an almost flat profile, with means between $1.171$
and $1.191$. The first covariance-weight update already produces
the increasing profile seen at $s=\star$; the remaining changes in
the odds-ratio means are at most $0.007$. The upward trend in these
point estimates should, however, be interpreted in light of uncertainty
about the age--smoking interaction.

\medskip{}

\begin{center}
{[}Table~\ref{tab:wheeze-or} here.{]}
\par\end{center}

\medskip{}

At all four ages, both the raw and sandwich-adjusted $95\%$ intervals
for the odds ratio include one. For example, at age $10$, the final
raw interval is $[0.854,1.960]$ and the adjusted interval is $[0.715,1.965]$.
The adjusted intervals are generally somewhat wider, especially at
the older ages, but neither construction excludes the null odds ratio
at any age. Thus, the increasing pattern in the quasi-posterior means
does not, by itself, establish an age-specific smoking association
or an increase in that association with age.

Overall, the application shows that replacing an arbitrary pilot weight
by a moment-covariance weight can materially change the quasi-posterior
center and uncertainty. The much smaller differences between $s=1$
and $s=\star$ indicate that self-consistency requires only a modest
further adjustment in this application.

\subsection{Sensitivity to moment rescaling}

\label{subsec:wheeze-ablation}

The application also allows us to examine the invariance of covariance
weighting to moment rescaling. We deliberately rescale the last moment
condition by considering
\[
\boldsymbol{D}_{\varphi}=\operatorname{diag}(1,\dots,1,\varphi),\qquad\boldsymbol{m}^{(\varphi)}_{i}(\boldsymbol{\theta})=\boldsymbol{D}_{\varphi}\boldsymbol{m}_{i}(\boldsymbol{\theta}),\qquad\varphi\in\{0.2,1,100\}.
\]
All other aspects of the analysis are unchanged. With a nonsingular
diagonal rescaling $\boldsymbol{D}$, it follows that 
\[
\widehat{\boldsymbol{C}}^{(\varphi)}_{N}(\boldsymbol{\vartheta})=\boldsymbol{D}_{\varphi}\widehat{\boldsymbol{C}}_{N}(\boldsymbol{\vartheta})\boldsymbol{D}^{\top}_{\varphi},
\]
so that
\[
\left\{ \widehat{\boldsymbol{C}}^{(\varphi)}_{N}(\boldsymbol{\vartheta})\right\} ^{-1}=\boldsymbol{D}^{-\top}_{\varphi}\widehat{\boldsymbol{C}}_{N}(\boldsymbol{\vartheta})^{-1}\boldsymbol{D}^{-1}_{\varphi}.
\]
Consequently,
\[
\bar{\boldsymbol{m}}^{(\varphi)}_{N}(\boldsymbol{\theta})^{\top}\left\{ \widehat{\boldsymbol{C}}^{(\varphi)}_{N}(\boldsymbol{\vartheta})\right\} ^{-1}\bar{\boldsymbol{m}}^{(\varphi)}_{N}(\boldsymbol{\theta})=\bar{\boldsymbol{m}}_{N}(\boldsymbol{\theta})^{\top}\widehat{\boldsymbol{C}}_{N}(\boldsymbol{\vartheta})^{-1}\bar{\boldsymbol{m}}_{N}(\boldsymbol{\theta}).
\]
This identity holds exactly for every candidate center $\boldsymbol{\vartheta}$;
hence, the ideal calibration map $\mathcal{T}_{N}(\boldsymbol{\vartheta})$
is invariant under such rescaling. Intermediate iterates may differ
because the identity-weight pilot produces different starting points.
The ideal fixed point is invariant, although numerical approximations
to it may differ because of Monte Carlo and numerical error. By contrast,
a fixed weight that does not adjust for the rescaling, such as the
identity weight used in the pilot, changes the relative contribution
of the moment conditions and can therefore affect the quasi-posterior,
particularly in the presence of an informative prior. Because the
pilot center itself depends on $\varphi$, intermediate CCQB iterates
need not be exactly identical across scalings. This sensitivity analysis
examines how rapidly CCQB removes this dependence on the initialization.

Table~\ref{tab:wheeze-ablation} focuses on $\theta_{2}$. At the
identity-weight pilot stage, the quasi-posterior mean varies with
the scaling of the last moment, taking values of $-0.147$, $-0.141$,
and $-0.064$ for $\varphi\in\{0.2,1,100\}$, respectively. The interval
summaries also exhibit sensitivity to rescaling: the pilot-stage sandwich-adjusted
$95\%$ interval excludes zero for $\varphi\in\{0.2,1\}$ but includes
zero for $\varphi=100$. After the first covariance-weight update,
the quasi-posterior means are all $-0.135$, and they remain between
$-0.136$ and $-0.135$ at $s=\star$. Both raw and sandwich-adjusted
intervals are also nearly identical across rescalings at these stages.
This marked reduction in sensitivity is consistent with the exact
rescaling invariance of the covariance-weighted criterion. Small remaining
numerical differences can be attributed to Monte Carlo approximation
and finite stopping tolerances.

\medskip{}

\begin{center}
{[}Table~\ref{tab:wheeze-ablation} here.{]}
\par\end{center}

\medskip{}

Additional parameter-wise sensitivity results are reported in the
Supplementary Material (Table~S9). They show that the identity-weight
pilot summaries depend on moment rescaling, whereas the quasi-posterior
means and intervals become nearly invariant after the first covariance-weight
update. Further iteration produces only minor changes.

\section{Conclusion}

\label{sec:conclusion}

This paper has proposed curvature-calibrated quasi-Bayesian (CCQB)
updating for moment-restricted models. In a quadratic moment quasi-posterior,
the weighting matrix determines the information scale as well as the
relative contribution of individual moments. CCQB uses the inverse
of an estimated moment covariance, or long-run covariance, evaluated
at a self-consistent quasi-posterior center while keeping the weight
fixed within each simulation run. The stopping rule distinguishes
changes in the calibrated center from the Monte Carlo error involved
in estimating them.

Under the stated local regularity conditions, the ideal calibration
map is contractive with probability approaching one, and its fixed
point is root-$N$ consistent. Given a Bernstein--von Mises approximation
at the population-optimal weight, the calibrated quasi-posterior and
its center have matching first-order covariance, equal to the inverse
Godambe information matrix. These results concern first-order calibration
under correctly specified, locally identified moment restrictions;
they do not guarantee exact finite-sample coverage.

The simulations indicate that the main improvement often comes from
replacing an arbitrary identity weight by a covariance weight. In
the overidentified IV design, one update and full iteration give similar
results. In the nonlinear Poisson design, further iteration has an
additional finite-sample effect: for $\theta_{1}$ it brings the coverage
of raw $95\%$ intervals closer to the nominal level, although undercoverage
persists for $\theta_{4}$. The supplementary dependent-data design
similarly demonstrates substantial benefits from covariance weighting
without uniformly nominal coverage. The sample-oracle benchmark often
performs better in smaller samples, and BETEL generally gives shorter
intervals at the cost of greater undercoverage there. The empirical
application illustrates how child-level moment covariance weighting
can materially change quasi-posterior means and uncertainty while
accommodating within-child dependence. It also confirms that further
updates after the first covariance weighting are small in this example.

Self-consistent iteration is therefore best viewed as a possible finite-sample
refinement of covariance weighting rather than as a requirement for
first-order validity or a guarantee of improved coverage. Important
extensions include misspecified moment models, regularized covariance
weighting when the moment dimension is large relative to the sample
size, and a systematic assessment of the computational cost of increasingly
stringent Monte Carlo tolerances.

\section*{Funding}

This work was supported by JSPS KAKENHI Grant Number 25K21168.

\bibliographystyle{apalike2}
\bibliography{reference}

\appendix
\makeatletter
\def\section@prefix{}
\def\@appsectionnostar#1{%
  \def\section@numbersep{.}%
  \old@section{#1}%
}
\makeatother

\part*{Appendix}

\section{Regularity conditions and proofs}

\label{sec:condition-proof}

This section presents the components of the main result separately.
We first state the local regularity conditions and an optional condition
that justifies pilot initialization. We then establish how the fixed-weight
quasi-posterior depends on its weighting matrix, use that result to
prove Theorem~\ref{thm:main-thm}, and derive bounds for iteration
and stopping errors from the contraction argument.

As in Section~\ref{sec:method}, all topological notions for subsets
of $\mathbb{S}^{K}$ are understood relative to $\mathbb{S}^{K}$.
Let 
\[
\boldsymbol{g}(\boldsymbol{\theta})=\mathbb{E}\left\{ \boldsymbol{m}_{i}(\boldsymbol{\theta})\right\} .
\]
For each fixed $\boldsymbol{W}\in\mathbb{S}^{K}_{++}$, let 
\[
\boldsymbol{\mu}_{N}(\boldsymbol{W})=\mathbb{E}_{\Pi^{\boldsymbol{W}}(\cdot\mid\mathcal{D}_{N})}(\boldsymbol{\theta})
\]
denote the mean of the fixed-weight quasi-posterior. For a fixed $\boldsymbol{W}$,
write 
\[
\mathbb{E}_{\boldsymbol{W}}(\cdot)=\mathbb{E}_{\Pi^{\boldsymbol{W}}(\cdot\mid\mathcal{D}_{N})}(\cdot),\qquad\operatorname{Cov}_{\boldsymbol{W}}(\cdot,\cdot)=\operatorname{Cov}_{\Pi^{\boldsymbol{W}}(\cdot\mid\mathcal{D}_{N})}(\cdot,\cdot).
\]

Recall the calibration map 
\[
\mathcal{T}_{N}(\boldsymbol{\vartheta})=\boldsymbol{\mu}_{N}\left\{ \boldsymbol{W}_{N}(\boldsymbol{\vartheta})\right\} ,
\]
whenever $\boldsymbol{W}_{N}(\boldsymbol{\vartheta})$ is well defined.
In the proof of Theorem~\ref{thm:main-thm}, we construct a fixed
neighborhood $\mathcal{U}_{0}$ of $\boldsymbol{\theta}_{0}$ on which
this map has the required properties with probability approaching
one.

For probability measures $P$ and $Q$ on $\mathbb{R}^{J}$, use the
conventions 
\[
d_{\mathrm{TV}}(P,Q)=\sup_{A}|P(A)-Q(A)|
\]
and 
\[
d_{\mathrm{BL}}(P,Q)=\sup_{\substack{\lVert f\rVert_{\infty}\leq1\\
\operatorname{Lip}(f)\leq1
}
}\left|\int f\,dP-\int f\,dQ\right|.
\]
For any fixed $\boldsymbol{W}\in\mathbb{S}^{K}_{++}$, define the
centered and scaled quasi-posterior law by 
\[
\mathsf{Q}^{\boldsymbol{W}}_{N}=\mathcal{L}_{\Pi^{\boldsymbol{W}}(\cdot\mid\mathcal{D}_{N})}\left\{ \sqrt{N}\left(\boldsymbol{\theta}-\boldsymbol{\mu}_{N}(\boldsymbol{W})\right)\right\} .
\]
The same notation is used when $\boldsymbol{W}$ is a $\sigma(\mathcal{D}_{N})$-measurable
random weighting matrix.

\subsection{Regularity conditions}

\label{subsec:condition}

\begin{assumption}

\label{ass:regularity-cond}

The following conditions hold. 
\begin{enumerate}
\item The function $\boldsymbol{g}(\boldsymbol{\theta})$ has $\boldsymbol{\theta}_{0}$
as its unique zero in a neighborhood of $\boldsymbol{\theta}_{0}$
and is continuously differentiable there. Differentiation and expectation
may be interchanged, so that 
\[
\boldsymbol{G}_{0}\coloneqq\frac{\partial\boldsymbol{g}(\boldsymbol{\theta}_{0})}{\partial\boldsymbol{\theta}^{\top}}=\mathbb{E}\left\{ \frac{\partial\boldsymbol{m}_{i}(\boldsymbol{\theta}_{0})}{\partial\boldsymbol{\theta}^{\top}}\right\} ,
\]
and $\boldsymbol{G}_{0}$ has full column rank. 
\item The matrix-valued function $\boldsymbol{C}(\boldsymbol{\theta})$
is continuously differentiable and positive definite on a neighborhood
$\mathcal{U}$ of $\boldsymbol{\theta}_{0}$, where $\boldsymbol{\theta}_{0}\in\operatorname{int}(\Theta)$.
For every $\boldsymbol{\theta}\in\mathcal{U}$, $\widehat{\boldsymbol{C}}_{N}(\boldsymbol{\theta})\in\mathbb{S}^{K}$,
and 
\[
\sup_{\boldsymbol{\theta}\in\mathcal{U}}\left\Vert \widehat{\boldsymbol{C}}_{N}(\boldsymbol{\theta})-\boldsymbol{C}(\boldsymbol{\theta})\right\Vert \xrightarrow{p}0,\qquad\inf_{\boldsymbol{\theta}\in\mathcal{U}}\lambda_{\min}\{\boldsymbol{C}(\boldsymbol{\theta})\}>0.
\]
\item The sample moments satisfy 
\[
\boldsymbol{\xi}_{N}=O_{p}(1).
\]
\item There exists a compact convex set $\mathcal{W}\subset\mathbb{S}^{K}_{++}$
such that $\boldsymbol{W}_{0}\in\operatorname{int}(\mathcal{W})$
and the following conditions hold. For each $\boldsymbol{W}\in\mathcal{W}$,
the fixed-weight quasi-posterior has a density $p^{\boldsymbol{W}}_{N}$
with respect to Lebesgue measure. For every $\boldsymbol{W},\boldsymbol{W}'\in\mathcal{W}$,
let 
\[
\boldsymbol{W}_{t}=(1-t)\boldsymbol{W}+t\boldsymbol{W}',\qquad0\leq t\leq1.
\]
The maps $t\mapsto\boldsymbol{\mu}_{N}(\boldsymbol{W}_{t})$ and $t\mapsto\sqrt{p^{\boldsymbol{W}_{t}}_{N}}$
are absolutely continuous, the latter as an $L^{2}$-valued map, and
their derivatives are obtained by differentiating under the integrals
defining the quasi-posterior normalizing constant and moments. These
functionals are jointly measurable in the data and the weighting matrix.
The following uniform quasi-posterior moment bounds hold: 
\[
R_{\theta,N}=\sup_{\boldsymbol{W}\in\mathcal{W}}N\mathbb{E}_{\boldsymbol{W}}\left[\lVert\boldsymbol{\theta}-\boldsymbol{\mu}_{N}(\boldsymbol{W})\rVert^{2}\right]=O_{p}(1)
\]
and 
\[
R_{m,N}\coloneqq\sup_{\boldsymbol{W}\in\mathcal{W}}N^{2}\mathbb{E}_{\boldsymbol{W}}\left\{ \lVert\bar{\boldsymbol{m}}_{N}(\boldsymbol{\theta})\rVert^{4}\right\} =O_{p}(1).
\]
In addition, 
\[
R_{\theta,4,N}=\sup_{\boldsymbol{W}\in\mathcal{W}}N^{2}\mathbb{E}_{\boldsymbol{W}}\left[\lVert\boldsymbol{\theta}-\boldsymbol{\mu}_{N}(\boldsymbol{W})\rVert^{4}\right]=O_{p}(1).
\]
\item At $\boldsymbol{W}_{0}=\boldsymbol{C}^{-1}_{0}$, the fixed-weight
quasi-posterior mean satisfies 
\[
\sqrt{N}\left\{ \boldsymbol{\mu}_{N}(\boldsymbol{W}_{0})-\boldsymbol{\theta}_{0}\right\} =-\boldsymbol{A}_{0}\boldsymbol{\xi}_{N}+o_{p}(1),
\]
and, conditionally on the data, 
\[
d_{\mathrm{BL}}\left[\mathsf{Q}^{\boldsymbol{W}_{0}}_{N},\;\mathcal{N}\left(\boldsymbol{0}_{J},\boldsymbol{V}_{0}\right)\right]\xrightarrow{p}0.
\]
\item The map $(\mathcal{D}_{N},\boldsymbol{\vartheta})\mapsto\boldsymbol{W}_{N}(\boldsymbol{\vartheta})$
is jointly measurable. There exists a measurable random variable $L_{N}=O_{p}(1)$
such that, with probability approaching one, the estimated weighting
map 
\[
\boldsymbol{W}_{N}(\boldsymbol{\vartheta})=\widehat{\boldsymbol{C}}_{N}(\boldsymbol{\vartheta})^{-1}
\]
is well defined on $\mathcal{U}$ and, uniformly for $\boldsymbol{\vartheta},\boldsymbol{\vartheta}'\in\mathcal{U}$,
\[
\lVert\boldsymbol{W}_{N}(\boldsymbol{\vartheta})-\boldsymbol{W}_{N}(\boldsymbol{\vartheta}')\rVert\leq L_{N}\lVert\boldsymbol{\vartheta}-\boldsymbol{\vartheta}'\rVert.
\]
\item The prior density is continuous and strictly positive in a neighborhood
of $\boldsymbol{\theta}_{0}$. 
\end{enumerate}
\end{assumption}

\subsection{Pilot initialization}

\label{subsec:pilot-init}

Let $\boldsymbol{W}^{\mathrm{p}}_{N}$ denote a possibly data-dependent
pilot weighting matrix that does not depend on the candidate calibration
center. A deterministic matrix, such as $\boldsymbol{I}_{K}$, is
included as a special case. Let $\mathcal{X}_{N}$ be the set of permitted
initial states for the pilot Markov chain, write $\widetilde{\boldsymbol{\mu}}^{\mathrm{p}}_{N,T_{\mathrm{p}}}(\boldsymbol{x})$
for the retained pilot-chain average when the chain is initialized
at $\boldsymbol{x}\in\mathcal{X}_{N}$, and let $\mathbb{P}^{\mathrm{p}}_{N,\boldsymbol{x}}(\cdot\mid\mathcal{D}_{N})$
denote the conditional law of the pilot simulation from that state.

\begin{assumption}[Pilot localization]

\label{ass:pilot-localization}

There exists a compact set $\mathcal{W}^{\mathrm{p}}\subset\mathbb{S}^{K}_{++}$
such that $\Pr\left\{ \boldsymbol{W}^{\mathrm{p}}_{N}\in\mathcal{W}^{\mathrm{p}}\right\} \to1$
and 
\[
\sup_{\boldsymbol{W}\in\mathcal{W}^{\mathrm{p}}}\lVert\boldsymbol{\mu}_{N}(\boldsymbol{W})-\boldsymbol{\theta}_{0}\rVert\xrightarrow{p}0.
\]
For every $\epsilon>0$, 
\[
\sup_{\boldsymbol{x}\in\mathcal{X}_{N}}\mathbb{P}^{\mathrm{p}}_{N,\boldsymbol{x}}\left\{ \left\Vert \widetilde{\boldsymbol{\mu}}^{\mathrm{p}}_{N,T_{\mathrm{p}}}(\boldsymbol{x})-\boldsymbol{\mu}_{N}(\boldsymbol{W}^{\mathrm{p}}_{N})\right\Vert >\epsilon\,\middle|\,\mathcal{D}_{N}\right\} \xrightarrow{p}0.
\]
\end{assumption}

The first condition requires global concentration of the pilot quasi-posterior;
it does not require the pilot weight to estimate $\boldsymbol{C}^{-1}_{0}$.
A standard route to sufficient conditions is global identification
and separation of the population moments: for every $\epsilon>0$,
\[
\inf_{\boldsymbol{\theta}\in\Theta:\,\lVert\boldsymbol{\theta}-\boldsymbol{\theta}_{0}\rVert\geq\epsilon}\lVert\boldsymbol{g}(\boldsymbol{\theta})\rVert>0,
\]
together with eigenvalues of matrices in $\mathcal{W}^{\mathrm{p}}$
that are bounded above and bounded away from zero, a uniform law of
large numbers for the fixed-weight criterion, local quadratic regularity
near $\boldsymbol{\theta}_{0}$, a prior that is positive near $\boldsymbol{\theta}_{0}$,
and either compactness of $\Theta$ or an appropriate posterior tail
and uniform-integrability condition. Such a uniform localization conclusion
generally need not hold when the population moments have multiple
separated zeros or, under misspecification, when the pseudo-true value
depends on the weighting matrix.

\begin{lemma}[Pilot initialization]

\label{lem:pilot-init}

Suppose that Assumption~\ref{ass:pilot-localization} holds and that
the pilot chain is initialized at any possibly random state $\boldsymbol{X}^{\mathrm{p}}_{N}\in\mathcal{X}_{N}$
that is either $\sigma(\mathcal{D}_{N})$-measurable or generated
by auxiliary randomization conditionally independent of the subsequent
pilot chain given $\mathcal{D}_{N}$. Then 
\[
\widetilde{\boldsymbol{\vartheta}}^{(0)}_{N}\coloneqq\widetilde{\boldsymbol{\mu}}^{\mathrm{p}}_{N,T_{\mathrm{p}}}\left(\boldsymbol{X}^{\mathrm{p}}_{N}\right)\xrightarrow{p}\boldsymbol{\theta}_{0}.
\]
Consequently, the pilot center lies in the contraction ball of Theorem~\ref{thm:main-thm}
with probability approaching one, and the center after one ideal calibration
update is first-order equivalent to the fixed point. If $\mathcal{X}_{N}=\Theta$,
the pilot chain may be initialized at an arbitrary state in the parameter
space. 

\end{lemma} 
\begin{proof}
By the triangle inequality, 
\[
\begin{aligned}\left\Vert \widetilde{\boldsymbol{\vartheta}}^{(0)}_{N}-\boldsymbol{\theta}_{0}\right\Vert  & \leq\left\Vert \widetilde{\boldsymbol{\mu}}^{\mathrm{p}}_{N,T_{\mathrm{p}}}\left(\boldsymbol{X}^{\mathrm{p}}_{N}\right)-\boldsymbol{\mu}_{N}\left(\boldsymbol{W}^{\mathrm{p}}_{N}\right)\right\Vert \\
 & \quad+\left\Vert \boldsymbol{\mu}_{N}\left(\boldsymbol{W}^{\mathrm{p}}_{N}\right)-\boldsymbol{\theta}_{0}\right\Vert =o_{p}(1).
\end{aligned}
\]
The remaining conclusions follow from Theorem~\ref{thm:main-thm}. 
\end{proof}

\subsection{Sensitivity to the weighting matrix}

\label{subsec:sensitivity}

The following lemma relates perturbations of the weighting matrix
to changes in both the quasi-posterior center and the centered quasi-posterior
law. It is used directly in the proof of Theorem~\ref{thm:main-thm}.

\begin{lemma}[Sensitivity with respect to the weighting matrix]

\label{lem:weight-sensitivity}

Suppose that Assumption~\ref{ass:regularity-cond}(4) holds. Then
\[
\kappa_{N}\coloneqq\sup_{\substack{\boldsymbol{W},\boldsymbol{W}'\in\mathcal{W}\\
\boldsymbol{W}\neq\boldsymbol{W}'
}
}\frac{\lVert\boldsymbol{\mu}_{N}(\boldsymbol{W})-\boldsymbol{\mu}_{N}(\boldsymbol{W}')\rVert}{\lVert\boldsymbol{W}-\boldsymbol{W}'\rVert}\leq\frac{\left(R_{\theta,N}R_{m,N}\right)^{1/2}}{2\sqrt{N}}.
\]
Consequently, 
\[
\kappa_{N}=O_{p}\left(N^{-1/2}\right).
\]
Moreover, 
\[
d_{\mathrm{TV}}\left(\Pi^{\boldsymbol{W}}(\cdot\mid\mathcal{D}_{N}),\;\Pi^{\boldsymbol{W}'}(\cdot\mid\mathcal{D}_{N})\right)\leq\frac{1}{4}R^{1/2}_{m,N}\lVert\boldsymbol{W}-\boldsymbol{W}'\rVert.
\]
Therefore, for every $\sigma(\mathcal{D}_{N})$-measurable sequence
$\widetilde{\boldsymbol{W}}_{N}$ taking values in $\mathcal{W}$
and satisfying 
\[
\left\Vert \widetilde{\boldsymbol{W}}_{N}-\boldsymbol{W}_{0}\right\Vert \xrightarrow{p}0,
\]
the centered quasi-posterior laws satisfy 
\[
d_{\mathrm{BL}}\left(\mathsf{Q}^{\widetilde{\boldsymbol{W}}_{N}}_{N},\mathsf{Q}^{\boldsymbol{W}_{0}}_{N}\right)\xrightarrow{p}0.
\]
\end{lemma} 
\begin{proof}
Fix $\boldsymbol{W},\boldsymbol{W}'\in\mathcal{W}$, set 
\[
\boldsymbol{H}=\boldsymbol{W}'-\boldsymbol{W},\qquad\boldsymbol{W}_{t}=(1-t)\boldsymbol{W}+t\boldsymbol{W}',
\]
and differentiate along this line segment. Define 
\[
q_{\boldsymbol{H}}(\boldsymbol{\theta})=\bar{\boldsymbol{m}}_{N}(\boldsymbol{\theta})^{\top}\boldsymbol{H}\bar{\boldsymbol{m}}_{N}(\boldsymbol{\theta}).
\]
Differentiating the normalized quasi-posterior in the direction $\boldsymbol{H}$
gives 
\[
D_{\boldsymbol{W}}\boldsymbol{\mu}_{N}(\boldsymbol{W})[\boldsymbol{H}]=-\frac{N}{2}\operatorname{Cov}_{\boldsymbol{W}}\{\boldsymbol{\theta},q_{\boldsymbol{H}}(\boldsymbol{\theta})\}.
\]
Since 
\[
\left|q_{\boldsymbol{H}}(\boldsymbol{\theta})\right|\leq\lVert\boldsymbol{H}\rVert\lVert\bar{\boldsymbol{m}}_{N}(\boldsymbol{\theta})\rVert^{2},
\]
the vector-valued Cauchy--Schwarz inequality yields 
\[
\begin{aligned}\lVert D_{\boldsymbol{W}}\boldsymbol{\mu}_{N}(\boldsymbol{W})[\boldsymbol{H}]\rVert & \leq\frac{N}{2}\left[\mathbb{E}_{\boldsymbol{W}}\left\{ \lVert\boldsymbol{\theta}-\boldsymbol{\mu}_{N}(\boldsymbol{W})\rVert^{2}\right\} \right]^{1/2}\\
 & \qquad\times\left[\operatorname{Var}_{\boldsymbol{W}}\{q_{\boldsymbol{H}}(\boldsymbol{\theta})\}\right]^{1/2}\\
 & \leq\frac{\left(R_{\theta,N}R_{m,N}\right)^{1/2}}{2\sqrt{N}}\lVert\boldsymbol{H}\rVert.
\end{aligned}
\]
For $\boldsymbol{H}=\boldsymbol{W}'-\boldsymbol{W}$, absolute continuity
along the line segment $\boldsymbol{W}_{t}=(1-t)\boldsymbol{W}+t\boldsymbol{W}'$
therefore implies 
\[
\lVert\boldsymbol{\mu}_{N}(\boldsymbol{W}')-\boldsymbol{\mu}_{N}(\boldsymbol{W})\rVert\leq\frac{\left(R_{\theta,N}R_{m,N}\right)^{1/2}}{2\sqrt{N}}\lVert\boldsymbol{W}'-\boldsymbol{W}\rVert.
\]
Taking the supremum proves the first assertion, and the assumed moment
bounds imply $\kappa_{N}=O_{p}(N^{-1/2})$.

For the law itself, differentiation of the square-root density gives
\[
D_{\boldsymbol{W}}\sqrt{p^{\boldsymbol{W}}_{N}(\boldsymbol{\theta})}[\boldsymbol{H}]=-\frac{N}{4}\sqrt{p^{\boldsymbol{W}}_{N}(\boldsymbol{\theta})}\left[q_{\boldsymbol{H}}(\boldsymbol{\theta})-\mathbb{E}_{\boldsymbol{W}}\left\{ q_{\boldsymbol{H}}(\boldsymbol{\theta})\right\} \right].
\]
Consequently, 
\[
\begin{aligned}\lVert D_{\boldsymbol{W}}\sqrt{p^{\boldsymbol{W}}_{N}}[\boldsymbol{H}]\rVert_{L^{2}} & =\frac{N}{4}\left[\operatorname{Var}_{\boldsymbol{W}}\left\{ q_{\boldsymbol{H}}(\boldsymbol{\theta})\right\} \right]^{1/2}\\
 & \leq\frac{1}{4}R^{1/2}_{m,N}\lVert\boldsymbol{H}\rVert.
\end{aligned}
\]
The $L^{2}$ absolute continuity in Assumption~\ref{ass:regularity-cond}(4)
therefore gives 
\[
\left\Vert \sqrt{p^{\boldsymbol{W}}_{N}}-\sqrt{p^{\boldsymbol{W}'}_{N}}\right\Vert _{L^{2}}\leq\frac{1}{4}R^{1/2}_{m,N}\lVert\boldsymbol{W}-\boldsymbol{W}'\rVert.
\]
Under the convention $d_{\mathrm{TV}}(P,Q)=\frac{1}{2}\int|p-q|$,
Cauchy--Schwarz gives 
\[
\begin{aligned}d_{\mathrm{TV}}(P,Q) & =\frac{1}{2}\int\left|\sqrt{p}-\sqrt{q}\right|\left|\sqrt{p}+\sqrt{q}\right|\\
 & \leq\frac{1}{2}\lVert\sqrt{p}-\sqrt{q}\rVert_{L^{2}}\lVert\sqrt{p}+\sqrt{q}\rVert_{L^{2}}\\
 & \leq\lVert\sqrt{p}-\sqrt{q}\rVert_{L^{2}}.
\end{aligned}
\]
This proves 
\[
d_{\mathrm{TV}}\left(\Pi^{\boldsymbol{W}}(\cdot\mid\mathcal{D}_{N}),\;\Pi^{\boldsymbol{W}'}(\cdot\mid\mathcal{D}_{N})\right)\leq\frac{1}{4}R^{1/2}_{m,N}\lVert\boldsymbol{W}-\boldsymbol{W}'\rVert.
\]

Finally, total variation is invariant under invertible affine transformations,
$d_{\mathrm{BL}}(P,Q)\leq2d_{\mathrm{TV}}(P,Q)$ under the conventions
above, and translation by $\boldsymbol{a}$ changes a law by at most
$\lVert\boldsymbol{a}\rVert$ in bounded-Lipschitz distance. Hence,
\[
\begin{aligned}d_{\mathrm{BL}}\left(\mathsf{Q}^{\boldsymbol{W}}_{N},\mathsf{Q}^{\boldsymbol{W}'}_{N}\right) & \leq2d_{\mathrm{TV}}\left(\Pi^{\boldsymbol{W}}(\cdot\mid\mathcal{D}_{N}),\;\Pi^{\boldsymbol{W}'}(\cdot\mid\mathcal{D}_{N})\right)\\
 & \quad+\sqrt{N}\lVert\boldsymbol{\mu}_{N}(\boldsymbol{W})-\boldsymbol{\mu}_{N}(\boldsymbol{W}')\rVert\\
 & \leq\frac{1}{2}\left\{ R^{1/2}_{m,N}+\left(R_{\theta,N}R_{m,N}\right)^{1/2}\right\} \lVert\boldsymbol{W}-\boldsymbol{W}'\rVert.
\end{aligned}
\]
The random coefficient in braces is $O_{p}(1)$. Taking $\boldsymbol{W}=\widetilde{\boldsymbol{W}}_{N}$
and $\boldsymbol{W}'=\boldsymbol{W}_{0}$ proves the final assertion. 
\end{proof}

The stochastic Lipschitz property of the fixed-weight quasi-posterior
mean and the local stability of the centered quasi-posterior law follow
from exact quasi-posterior sensitivity identities and the uniform
moment bounds in Assumption~\ref{ass:regularity-cond}(4), as shown
in Lemma~\ref{lem:weight-sensitivity}. Thus, these properties need
not be imposed separately. The first-order mean expansion and the
Bernstein--von Mises approximation are imposed only at the efficient
population weight $\boldsymbol{W}_{0}$.

\subsection{Proof of Theorem~\ref{thm:main-thm}}

\label{subsec:proof-main-thm}

The proof proceeds in four steps: we establish a local Lipschitz bound
for the calibration map, show that it maps a fixed neighborhood into
itself and is contractive there, derive the fixed-point expansion
and weight consistency, and show that the Gaussian approximation and
second-moment convergence continue to hold under the calibrated random
weight.
\begin{proof}
Define 
\[
\mathcal{T}_{N}(\boldsymbol{\vartheta})=\boldsymbol{\mu}_{N}\{\boldsymbol{W}_{N}(\boldsymbol{\vartheta})\}.
\]
By Lemma~\ref{lem:weight-sensitivity}, 
\begin{equation}
\lVert\boldsymbol{\mu}_{N}(\boldsymbol{W})-\boldsymbol{\mu}_{N}(\boldsymbol{W}')\rVert\leq\kappa_{N}\lVert\boldsymbol{W}-\boldsymbol{W}'\rVert,\qquad\kappa_{N}=O_{p}\left(N^{-1/2}\right).\label{eq:posterior-mean-sensitivity}
\end{equation}

Write 
\[
\boldsymbol{W}(\boldsymbol{\vartheta})=\boldsymbol{C}(\boldsymbol{\vartheta})^{-1}.
\]
Because $\boldsymbol{W}(\boldsymbol{\theta}_{0})=\boldsymbol{W}_{0}\in\operatorname{int}(\mathcal{W})$
and $\boldsymbol{W}(\cdot)$ is continuous at $\boldsymbol{\theta}_{0}$,
there exist a neighborhood $\mathcal{U}_{0}$ of $\boldsymbol{\theta}_{0}$,
with compact closure contained in $\mathcal{U}\cap\operatorname{int}(\Theta)$,
and a constant $\varepsilon>0$ such that 
\[
\inf_{\boldsymbol{\vartheta}\in\overline{\mathcal{U}}_{0}}\operatorname{dist}_{\mathbb{S}^{K}}\left\{ \boldsymbol{W}(\boldsymbol{\vartheta}),\;\mathbb{S}^{K}\setminus\mathcal{W}\right\} \geq2\varepsilon.
\]
The uniform lower eigenvalue bound for $\boldsymbol{C}(\boldsymbol{\vartheta})$,
the uniform consistency of $\widehat{\boldsymbol{C}}_{N}(\boldsymbol{\vartheta})$,
and the pointwise identity 
\[
\widehat{\boldsymbol{C}}_{N}(\boldsymbol{\vartheta})^{-1}-\boldsymbol{C}(\boldsymbol{\vartheta})^{-1}=\widehat{\boldsymbol{C}}_{N}(\boldsymbol{\vartheta})^{-1}\left\{ \boldsymbol{C}(\boldsymbol{\vartheta})-\widehat{\boldsymbol{C}}_{N}(\boldsymbol{\vartheta})\right\} \boldsymbol{C}(\boldsymbol{\vartheta})^{-1}
\]
imply 
\[
\sup_{\boldsymbol{\vartheta}\in\mathcal{U}_{0}}\lVert\boldsymbol{W}_{N}(\boldsymbol{\vartheta})-\boldsymbol{W}(\boldsymbol{\vartheta})\rVert=o_{p}(1).
\]
Consequently, 
\begin{equation}
\Pr\left\{ \boldsymbol{W}_{N}(\boldsymbol{\vartheta})\in\mathcal{W}\text{ for every }\boldsymbol{\vartheta}\in\mathcal{U}_{0}\right\} \to1.\label{eq:weight-range-event}
\end{equation}

On the event in \eqref{eq:weight-range-event}, for every $\boldsymbol{\vartheta},\boldsymbol{\vartheta}'\in\mathcal{U}_{0}$,
\eqref{eq:posterior-mean-sensitivity} and Assumption~\ref{ass:regularity-cond}(6)
give 
\[
\begin{aligned}\lVert\mathcal{T}_{N}(\boldsymbol{\vartheta})-\mathcal{T}_{N}(\boldsymbol{\vartheta}')\rVert & \leq\kappa_{N}\lVert\boldsymbol{W}_{N}(\boldsymbol{\vartheta})-\boldsymbol{W}_{N}(\boldsymbol{\vartheta}')\rVert\\
 & \leq\kappa_{N}L_{N}\lVert\boldsymbol{\vartheta}-\boldsymbol{\vartheta}'\rVert.
\end{aligned}
\]
Set 
\[
\rho_{N}=\kappa_{N}L_{N}.
\]
Then 
\[
\rho_{N}=O_{p}\left(N^{-1/2}\right)\qquad\text{and}\qquad\Pr(\rho_{N}<1)\to1.
\]

We next establish that the calibration map is a self-map. On the event
in \eqref{eq:weight-range-event}, 
\[
\begin{aligned}\sup_{\boldsymbol{\vartheta}\in\mathcal{U}_{0}}\lVert\mathcal{T}_{N}(\boldsymbol{\vartheta})-\boldsymbol{\theta}_{0}\rVert & \leq\sup_{\boldsymbol{\vartheta}\in\mathcal{U}_{0}}\lVert\boldsymbol{\mu}_{N}\{\boldsymbol{W}_{N}(\boldsymbol{\vartheta})\}-\boldsymbol{\mu}_{N}(\boldsymbol{W}_{0})\rVert\\
 & \quad+\lVert\boldsymbol{\mu}_{N}(\boldsymbol{W}_{0})-\boldsymbol{\theta}_{0}\rVert\\
 & \leq\kappa_{N}\operatorname{diam}(\mathcal{W})+\lVert\boldsymbol{\mu}_{N}(\boldsymbol{W}_{0})-\boldsymbol{\theta}_{0}\rVert.
\end{aligned}
\]
Assumptions~\ref{ass:regularity-cond}(3) and \ref{ass:regularity-cond}(5)
imply 
\[
\sqrt{N}\{\boldsymbol{\mu}_{N}(\boldsymbol{W}_{0})-\boldsymbol{\theta}_{0}\}=-\boldsymbol{A}_{0}\boldsymbol{\xi}_{N}+o_{p}(1)=O_{p}(1).
\]
Since $\mathcal{W}$ is compact, 
\begin{equation}
\sup_{\boldsymbol{\vartheta}\in\mathcal{U}_{0}}\lVert\mathcal{T}_{N}(\boldsymbol{\vartheta})-\boldsymbol{\theta}_{0}\rVert=O_{p}(N^{-1/2}).\label{eq:uniform-map-bound}
\end{equation}

Choose a fixed $\delta>0$ such that 
\[
\overline{\mathcal{B}}_{\delta}(\boldsymbol{\theta}_{0})\subset\mathcal{U}_{0}.
\]
Equations~\eqref{eq:weight-range-event} and \eqref{eq:uniform-map-bound},
together with $\Pr(\rho_{N}<1)\to1$, imply that there is an event
$\mathcal{E}_{N}$ with $\Pr(\mathcal{E}_{N})\to1$ on which $\mathcal{T}_{N}$
maps $\overline{\mathcal{B}}_{\delta}(\boldsymbol{\theta}_{0})$ into
itself and is a contraction there with modulus $\rho_{N}$. The Banach
fixed-point theorem therefore gives a unique fixed point $\boldsymbol{\vartheta}^{\star}_{N}$
in this ball on $\mathcal{E}_{N}$, and the Picard iteration converges
to it from every starting point in the ball. Hence, any possibly random
sequence satisfying 
\[
\Pr\left\{ \boldsymbol{\vartheta}^{(0)}_{N}\in\overline{\mathcal{B}}_{\delta}(\boldsymbol{\theta}_{0})\right\} \to1
\]
converges to the fixed point with probability approaching one. On
the intersection of this initialization event and $\mathcal{E}_{N}$,
\eqref{eq:uniform-map-bound} gives 
\[
\boldsymbol{\vartheta}^{(1)}_{N}-\boldsymbol{\theta}_{0}=O_{p}\left(N^{-1/2}\right),
\]
whereas the contraction inequality gives 
\[
\left\Vert \boldsymbol{\vartheta}^{(1)}_{N}-\boldsymbol{\vartheta}^{\star}_{N}\right\Vert \leq2\delta\rho_{N}=O_{p}\left(N^{-1/2}\right)
\]
and 
\[
\left\Vert \boldsymbol{\vartheta}^{(2)}_{N}-\boldsymbol{\vartheta}^{\star}_{N}\right\Vert \leq2\delta\rho^{2}_{N}=O_{p}\left(N^{-1}\right).
\]
If $\boldsymbol{\vartheta}^{(0)}_{N}\xrightarrow{p}\boldsymbol{\theta}_{0}$,
then $\boldsymbol{\vartheta}^{(0)}_{N}-\boldsymbol{\vartheta}^{\star}_{N}=o_{p}(1)$,
and therefore 
\[
\boldsymbol{\vartheta}^{(1)}_{N}-\boldsymbol{\vartheta}^{\star}_{N}=o_{p}\left(N^{-1/2}\right).
\]

By the joint measurability conditions in Assumption~\ref{ass:regularity-cond}(4)
and \ref{ass:regularity-cond}(6), the fixed point on $\mathcal{E}_{N}$
can be selected measurably as the limit of Picard iterates initialized
at the deterministic point $\boldsymbol{\theta}_{0}$. Define the
following extensions: 
\[
\boldsymbol{\vartheta}^{\star}_{N}=\boldsymbol{\mu}_{N}(\boldsymbol{W}_{0})\quad\text{on }\mathcal{E}^{c}_{N},
\]
and 
\[
\boldsymbol{W}^{\star}_{N}=\begin{cases}
\boldsymbol{W}_{N}\left(\boldsymbol{\vartheta}^{\star}_{N}\right), & \text{on }\mathcal{E}_{N},\\
\boldsymbol{W}_{0}, & \text{on }\mathcal{E}^{c}_{N}.
\end{cases}
\]
Then, for every sample outcome, 
\begin{equation}
\boldsymbol{\vartheta}^{\star}_{N}=\boldsymbol{\mu}_{N}\left(\boldsymbol{W}^{\star}_{N}\right).\label{eq:extended-fixed-point}
\end{equation}

On $\mathcal{E}_{N}$, the fixed-point property and \eqref{eq:uniform-map-bound}
imply 
\[
\left\Vert \boldsymbol{\vartheta}^{\star}_{N}-\boldsymbol{\theta}_{0}\right\Vert \leq\sup_{\boldsymbol{\vartheta}\in\mathcal{U}_{0}}\lVert\mathcal{T}_{N}(\boldsymbol{\vartheta})-\boldsymbol{\theta}_{0}\rVert.
\]
On $\mathcal{E}^{c}_{N}$, the same rate follows from the expansion
of $\boldsymbol{\mu}_{N}(\boldsymbol{W}_{0})$. Hence, 
\[
\boldsymbol{\vartheta}^{\star}_{N}-\boldsymbol{\theta}_{0}=O_{p}\left(N^{-1/2}\right).
\]

The uniform inverse consistency established above and the continuity
of $\boldsymbol{W}(\cdot)$ give 
\[
\begin{aligned}\mathbf{1}\{\mathcal{E}_{N}\}\left\Vert \boldsymbol{W}^{\star}_{N}-\boldsymbol{W}_{0}\right\Vert  & \leq\sup_{\boldsymbol{\vartheta}\in\mathcal{U}_{0}}\left\Vert \boldsymbol{W}_{N}(\boldsymbol{\vartheta})-\boldsymbol{W}(\boldsymbol{\vartheta})\right\Vert \\
 & \quad+\left\Vert \boldsymbol{W}\left(\boldsymbol{\vartheta}^{\star}_{N}\right)-\boldsymbol{W}_{0}\right\Vert =o_{p}(1).
\end{aligned}
\]
Since $\boldsymbol{W}^{\star}_{N}=\boldsymbol{W}_{0}$ on $\mathcal{E}^{c}_{N}$,
\begin{equation}
\boldsymbol{W}^{\star}_{N}\xrightarrow{p}\boldsymbol{W}_{0}.\label{eq:calibrated-weight-consistency}
\end{equation}
Because $\boldsymbol{W}^{\star}_{N}$ agrees with $\boldsymbol{W}_{N}(\boldsymbol{\vartheta}^{\star}_{N})$
with probability approaching one, this proves the asserted consistency
for any measurable extension of the latter on $\mathcal{E}^{c}_{N}$.

By \eqref{eq:posterior-mean-sensitivity}, \eqref{eq:calibrated-weight-consistency},
and $\sqrt{N}\kappa_{N}=O_{p}(1)$, 
\[
\sqrt{N}\left\Vert \boldsymbol{\mu}_{N}\left(\boldsymbol{W}^{\star}_{N}\right)-\boldsymbol{\mu}_{N}(\boldsymbol{W}_{0})\right\Vert \leq\sqrt{N}\kappa_{N}\left\Vert \boldsymbol{W}^{\star}_{N}-\boldsymbol{W}_{0}\right\Vert =o_{p}(1).
\]
Combining this result with \eqref{eq:extended-fixed-point} and the
mean expansion in Assumption~\ref{ass:regularity-cond}(5) yields
\[
\sqrt{N}\left(\boldsymbol{\vartheta}^{\star}_{N}-\boldsymbol{\theta}_{0}\right)=-\boldsymbol{A}_{0}\boldsymbol{\xi}_{N}+o_{p}(1).
\]

It remains to establish the conditional Gaussian approximation. The
matrix $\boldsymbol{W}^{\star}_{N}$ is $\sigma(\mathcal{D}_{N})$-measurable,
takes values in $\mathcal{W}$, and converges in probability to $\boldsymbol{W}_{0}$.
Lemma~\ref{lem:weight-sensitivity} therefore implies 
\[
d_{\mathrm{BL}}\left(\mathsf{Q}^{\boldsymbol{W}^{\star}_{N}}_{N},\mathsf{Q}^{\boldsymbol{W}_{0}}_{N}\right)\xrightarrow{p}0.
\]
The Bernstein--von Mises condition in Assumption~\ref{ass:regularity-cond}(5)
and the triangle inequality then give 
\[
d_{\mathrm{BL}}\left[\mathsf{Q}^{\boldsymbol{W}^{\star}_{N}}_{N},\;\mathcal{N}\left(\boldsymbol{0}_{J},\boldsymbol{V}_{0}\right)\right]\xrightarrow{p}0.
\]
Using \eqref{eq:extended-fixed-point}, this is 
\[
d_{\mathrm{BL}}\left[\mathcal{L}_{\Pi^{\boldsymbol{W}^{\star}_{N}}(\cdot\mid\mathcal{D}_{N})}\left\{ \sqrt{N}\left(\boldsymbol{\theta}-\boldsymbol{\vartheta}^{\star}_{N}\right)\right\} ,\;\mathcal{N}\left(\boldsymbol{0}_{J},\boldsymbol{V}_{0}\right)\right]\xrightarrow{p}0.
\]
On $\mathcal{E}_{N}$, $\boldsymbol{W}^{\star}_{N}=\boldsymbol{W}_{N}\left(\boldsymbol{\vartheta}^{\star}_{N}\right)$;
hence the display is equivalent to the asserted calibrated quasi-posterior
approximation.

If, in addition,
\[
\boldsymbol{\xi}_{N}=\sqrt{N}\bar{\boldsymbol{m}}_{N}(\boldsymbol{\theta}_{0})\rightsquigarrow\mathcal{N}\left(\boldsymbol{0}_{K},\boldsymbol{C}_{0}\right),
\]
then the fixed-point expansion and Slutsky's theorem imply 
\[
\sqrt{N}\left(\boldsymbol{\vartheta}^{\star}_{N}-\boldsymbol{\theta}_{0}\right)\rightsquigarrow\mathcal{N}\left(\boldsymbol{0}_{J},\boldsymbol{A}_{0}\boldsymbol{C}_{0}\boldsymbol{A}^{\top}_{0}\right).
\]
Since 
\[
\boldsymbol{A}_{0}\boldsymbol{C}_{0}\boldsymbol{A}^{\top}_{0}=\left(\boldsymbol{G}^{\top}_{0}\boldsymbol{C}^{-1}_{0}\boldsymbol{G}_{0}\right)^{-1}=\boldsymbol{V}_{0},
\]
the stated asymptotic normality follows. 

Finally, we use the Gaussian approximation to establish convergence
of the quasi-posterior covariance. Let $X_{N}=\sqrt{N}\left(\boldsymbol{\theta}-\boldsymbol{\vartheta}^{\star}_{N}\right)$
under the calibrated quasi-posterior. Assumption~\ref{ass:regularity-cond}(4)
implies
\[
\mathbb{E}_{\Pi^{\boldsymbol{W}^{\star}_{N}}}\lVert X_{N}\rVert^{4}\leq R_{\theta,4,N}=O_{p}(1).
\]
For each entry $X_{N,j}X_{N,k}$, truncate the corresponding function
outside a ball of radius $M$. Bounded-Lipschitz convergence implies
convergence of the expectation of the truncated function, while the
fourth-moment bound controls the truncation remainder by $O_{p}\left(M^{-2}\right)$.
Letting first $N\to\infty$ and then $M\to\infty$ yields convergence
in probability of the second-moment matrices. Since the centered quasi-posterior
has mean zero, this proves
\[
N\operatorname{Var}_{\Pi^{\boldsymbol{W}^{\star}_{N}}}(\boldsymbol{\theta})\xrightarrow{p}\boldsymbol{V}_{0}.
\]
\end{proof}

\subsection{Iteration error and stopping}

\label{subsec:iter-stop}

Theorem~\ref{thm:main-thm} gives the iteration rates for locally
admissible and consistent initial centers. Here we derive an a posteriori
stopping-error bound and quantify the additional error from evaluating
the calibration map by Monte Carlo.

Suppose that the ideal iteration is stopped at an index $S$ satisfying
\[
\left\Vert \boldsymbol{\vartheta}^{(S)}_{N}-\boldsymbol{\vartheta}^{(S-1)}_{N}\right\Vert \leq\epsilon_{N}.
\]
On the contraction event, the standard a posteriori error bound gives
\[
\begin{aligned}\left\Vert \boldsymbol{\vartheta}^{(S)}_{N}-\boldsymbol{\vartheta}^{\star}_{N}\right\Vert  & \leq\frac{\rho_{N}}{1-\rho_{N}}\left\Vert \boldsymbol{\vartheta}^{(S)}_{N}-\boldsymbol{\vartheta}^{(S-1)}_{N}\right\Vert \\
 & \leq\frac{\rho_{N}\epsilon_{N}}{1-\rho_{N}}.
\end{aligned}
\]
Because $\rho_{N}=O_{p}\left(N^{-1/2}\right)$, the stopped ideal
iterate is first-order equivalent to the fixed point for any deterministic
tolerance $\epsilon_{N}=o(1)$, and more generally for any random
tolerance $\epsilon_{N}=o_{p}(1)$.

\begin{remark}[Monte Carlo approximation]

The pilot and calibration runs have different Monte Carlo accuracy
requirements. The pilot run is used only to enter the local contraction
region. Accordingly, it is enough that 
\[
\widetilde{\boldsymbol{\mu}}^{\mathrm{p}}_{N,T_{\mathrm{p}}}-\boldsymbol{\mu}_{N}\left(\boldsymbol{W}^{\mathrm{p}}_{N}\right)=o_{p}(1),
\]
uniformly over the permitted pilot-chain initial states when arbitrary
initialization is claimed. By contrast, the calibration runs determine
the final center on the statistical scale $N^{-1/2}$.

Write the practical calibration update as an inexact Picard iteration,
\[
\widetilde{\boldsymbol{\vartheta}}^{(s+1)}_{N}=\mathcal{T}_{N}\left(\widetilde{\boldsymbol{\vartheta}}^{(s)}_{N}\right)+\boldsymbol{e}^{(s)}_{N},
\]
and suppose that, over the executed calibration runs, 
\[
\delta_{N}\coloneqq\sup_{s}\left\Vert \boldsymbol{e}^{(s)}_{N}\right\Vert =o_{p}\left(N^{-1/2}\right).
\]
On the contraction event, if the practical iteration stops at $S$,
then 
\[
\left\Vert \widetilde{\boldsymbol{\vartheta}}^{(S)}_{N}-\boldsymbol{\vartheta}^{\star}_{N}\right\Vert \leq\frac{\rho_{N}\left\Vert \widetilde{\boldsymbol{\vartheta}}^{(S)}_{N}-\widetilde{\boldsymbol{\vartheta}}^{(S-1)}_{N}\right\Vert +\delta_{N}}{1-\rho_{N}}.
\]
Assume further that the eigenvalues of $N\widehat{\boldsymbol{\Sigma}}^{\mathrm{ref}}_{N}$
are bounded above and bounded away from zero in probability. A fixed
standardized stopping threshold $\tau$ then implies 
\[
\left\Vert \widetilde{\boldsymbol{\vartheta}}^{(S)}_{N}-\widetilde{\boldsymbol{\vartheta}}^{(S-1)}_{N}\right\Vert =O_{p}\left(N^{-1/2}\right).
\]
Consequently, the stopped practical iterate is first-order equivalent
to $\boldsymbol{\vartheta}^{\star}_{N}$ whenever 
\[
\delta_{N}=o_{p}\left(N^{-1/2}\right).
\]
For example, suppose that the number of executed calibration runs
is stochastically bounded and that a Markov-chain central limit theorem
holds uniformly over those runs, with chain-mean asymptotic variances
of order $N^{-1}$. If the minimum effective sample size over the
executed runs diverges in probability, then 
\[
\delta_{N}=o_{p}\left(N^{-1/2}\right).
\]

\end{remark}

To relate the stopped iterate to the final retained target, suppose
that

\[
\left\Vert \widetilde{\boldsymbol{\vartheta}}^{(S)}_{N}-\boldsymbol{\vartheta}^{\star}_{N}\right\Vert =o_{p}\left(N^{-1/2}\right).
\]
Then
\[
\left\Vert \boldsymbol{W}_{N}\left(\widetilde{\boldsymbol{\vartheta}}^{(S)}_{N}\right)-\boldsymbol{W}^{\star}_{N}\right\Vert =o_{p}\left(N^{-1/2}\right),
\]
and, using $\kappa_{N}=O_{p}\left(N^{-1/2}\right)$,
\[
\left\Vert \boldsymbol{\mu}_{N}\left\{ \boldsymbol{W}_{N}\left(\widetilde{\boldsymbol{\vartheta}}^{(S)}_{N}\right)\right\} -\boldsymbol{\vartheta}^{\star}_{N}\right\Vert =o_{p}\left(N^{-1}\right).
\]
Thus, the final retained quasi-posterior has the same first-order
Gaussian limit as the ideal calibrated target.

\setcounter{table}{0}
\renewcommand{\thetable}{\arabic{table}}

\clearpage{}

\begin{table}
\caption{Number of CCQB calibration updates}

\label{tab:updates}

\medskip{}

\centering{}%
\begin{tabular}{lrrrrr}
\hline 
DGP & $N$ & \multicolumn{3}{c}{\begin{cellvarwidth}[t]
\raggedleft
Number of CCQB

calibration updates
\end{cellvarwidth}} & \multirow{2}{*}{\begin{cellvarwidth}[t]
\raggedleft
Number of

failed runs
\end{cellvarwidth}}\tabularnewline
\cline{3-5}
 &  & Min. & Median & Max. & \tabularnewline
\hline 
\multirow{4}{*}{IV} & 50 & 1 & 3 & 8 & 0\tabularnewline
 & 100 & 1 & 2 & 4 & 0\tabularnewline
 & 200 & 1 & 2 & 3 & 0\tabularnewline
 & 500 & 1 & 2 & 2 & 0\tabularnewline
\hline 
\multirow{4}{*}{Poisson} & 50 & 2 & 3 & 7 & 0\tabularnewline
 & 100 & 2 & 3 & 6 & 0\tabularnewline
 & 200 & 2 & 3 & 6 & 0\tabularnewline
 & 500 & 2 & 3 & 4 & 0\tabularnewline
\hline 
\end{tabular}
\end{table}

\clearpage{}

\begin{table}
\caption{$95\%$ nominal coverage of $\theta_{1}$}

\label{tab:coverage}

\medskip{}

\centering{}%
\begin{tabular}{llccrrrr}
\hline 
DGP & Method & $s$ & Cov. & \multicolumn{4}{c}{$N$}\tabularnewline
\cline{5-8}
 &  &  &  & 50 & 100 & 200 & 500\tabularnewline
\hline 
\multirow{9}{*}{IV} & \multirow{6}{*}{CCQB} & \multirow{2}{*}{$0$} & raw & 0.342 & 0.326 & 0.349 & 0.361\tabularnewline
 &  &  & adj & 0.913 & 0.913 & 0.934 & 0.960\tabularnewline
\cline{3-8}
 &  & \multirow{2}{*}{$1$} & raw & 0.873 & 0.887 & 0.923 & 0.957\tabularnewline
 &  &  & adj & 0.875 & 0.886 & 0.923 & 0.957\tabularnewline
\cline{3-8}
 &  & \multirow{2}{*}{$\star$} & raw & 0.865 & 0.883 & 0.922 & 0.957\tabularnewline
 &  &  & adj & 0.866 & 0.883 & 0.923 & 0.957\tabularnewline
\cline{2-8}
 & \multirow{2}{*}{Oracle} & \multirow{2}{*}{} & raw & 0.922 & 0.918 & 0.935 & 0.959\tabularnewline
 &  &  & adj & 0.898 & 0.900 & 0.923 & 0.957\tabularnewline
\cline{2-8}
 & BETEL &  &  & 0.807 & 0.878 & 0.908 & 0.951\tabularnewline
\hline 
\multirow{9}{*}{Poisson} & \multirow{6}{*}{CCQB} & \multirow{2}{*}{$0$} & raw & 1.000 & 1.000 & 1.000 & 1.000\tabularnewline
 &  &  & adj & 1.000 & 1.000 & 0.995 & 0.939\tabularnewline
\cline{3-8}
 &  & \multirow{2}{*}{$1$} & raw & 0.986 & 0.989 & 0.986 & 0.986\tabularnewline
 &  &  & adj & 0.976 & 0.973 & 0.956 & 0.958\tabularnewline
\cline{3-8}
 &  & \multirow{2}{*}{$\star$} & raw & 0.925 & 0.937 & 0.942 & 0.958\tabularnewline
 &  &  & adj & 0.943 & 0.941 & 0.949 & 0.960\tabularnewline
\cline{2-8}
 & \multirow{2}{*}{Oracle} &  & raw & 0.961 & 0.958 & 0.954 & 0.961\tabularnewline
 &  &  & adj & 0.949 & 0.943 & 0.943 & 0.959\tabularnewline
\cline{2-8}
 & BETEL &  &  & 0.900 & 0.918 & 0.934 & 0.949\tabularnewline
\hline 
\end{tabular}
\end{table}

\clearpage{}

\begin{table}
\caption{Covariance calibration}

\label{tab:cov-calib}

\medskip{}

\centering{}%
\begin{tabular}{llccrrrr}
\hline 
DGP & Method & $s$ & Cov. & \multicolumn{4}{c}{$N$}\tabularnewline
\cline{5-8}
 &  &  &  & 50 & 100 & 200 & 500\tabularnewline
\hline 
\multirow{9}{*}{IV} & \multirow{6}{*}{CCQB} & \multirow{2}{*}{$0$} & raw & 3.141 & 3.026 & 2.984 & 2.884\tabularnewline
 &  &  & adj & 0.657 & 0.439 & 0.310 & 0.195\tabularnewline
\cline{3-8}
 &  & \multirow{2}{*}{$1$} & raw & 0.738 & 0.470 & 0.320 & 0.193\tabularnewline
 &  &  & adj & 0.743 & 0.472 & 0.318 & 0.192\tabularnewline
\cline{3-8}
 &  & \multirow{2}{*}{$\star$} & raw & 0.780 & 0.480 & 0.320 & 0.192\tabularnewline
 &  &  & adj & 0.781 & 0.478 & 0.320 & 0.192\tabularnewline
\cline{2-8}
 & \multirow{2}{*}{Oracle} & \multirow{2}{*}{} & raw & 0.612 & 0.422 & 0.297 & 0.191\tabularnewline
 &  &  & adj & 0.665 & 0.441 & 0.306 & 0.194\tabularnewline
\cline{2-8}
 & BETEL &  &  & 0.974 & 0.573 & 0.368 & 0.209\tabularnewline
\hline 
\multirow{9}{*}{Poisson} & \multirow{6}{*}{CCQB} & \multirow{2}{*}{$0$} & raw & 4.222 & 4.192 & 4.093 & 4.030\tabularnewline
 &  &  & adj & 3.278 & 2.719 & 2.120 & 1.295\tabularnewline
\cline{3-8}
 &  & \multirow{2}{*}{$1$} & raw & 0.964 & 0.923 & 0.853 & 0.689\tabularnewline
 &  &  & adj & 0.884 & 0.615 & 0.442 & 0.310\tabularnewline
\cline{3-8}
 &  & \multirow{2}{*}{$\star$} & raw & 0.862 & 0.638 & 0.440 & 0.306\tabularnewline
 &  &  & adj & 0.941 & 0.663 & 0.452 & 0.310\tabularnewline
\cline{2-8}
 & \multirow{2}{*}{Oracle} &  & raw & 0.724 & 0.557 & 0.429 & 0.304\tabularnewline
 &  &  & adj & 0.852 & 0.614 & 0.443 & 0.310\tabularnewline
\cline{2-8}
 & BETEL &  &  & 0.780 & 0.577 & 0.437 & 0.297\tabularnewline
\hline 
\end{tabular}
\end{table}

\clearpage{}

\begin{table}
\caption{Mean lengths of nominal $95\%$ intervals for $\theta_{1}$}

\label{tab:length}

\medskip{}

\centering{}%
\begin{tabular}{llccrrrr}
\hline 
DGP & Method & $s$ & Cov. & \multicolumn{4}{c}{$N$}\tabularnewline
\cline{5-8}
 &  &  &  & 50 & 100 & 200 & 500\tabularnewline
\hline 
\multirow{9}{*}{IV} & \multirow{6}{*}{CCQB} & \multirow{2}{*}{$0$} & raw & 0.498 & 0.334 & 0.233 & 0.145\tabularnewline
 &  &  & adj & 2.013 & 1.397 & 0.993 & 0.627\tabularnewline
\cline{3-8}
 &  & \multirow{2}{*}{$1$} & raw & 1.786 & 1.315 & 0.961 & 0.619\tabularnewline
 &  &  & adj & 1.796 & 1.319 & 0.962 & 0.620\tabularnewline
\cline{3-8}
 &  & \multirow{2}{*}{$\star$} & raw & 1.772 & 1.312 & 0.960 & 0.619\tabularnewline
 &  &  & adj & 1.790 & 1.318 & 0.962 & 0.620\tabularnewline
\cline{2-8}
 & \multirow{2}{*}{Oracle} & \multirow{2}{*}{} & raw & 1.872 & 1.350 & 0.973 & 0.622\tabularnewline
 &  &  & adj & 1.819 & 1.324 & 0.963 & 0.620\tabularnewline
\cline{2-8}
 & BETEL &  &  & 1.641 & 1.244 & 0.935 & 0.613\tabularnewline
\hline 
\multirow{9}{*}{Poisson} & \multirow{6}{*}{CCQB} & \multirow{2}{*}{$0$} & raw & 16.065 & 14.670 & 13.094 & 10.704\tabularnewline
 &  &  & adj & 9.748 & 6.711 & 4.528 & 2.480\tabularnewline
\cline{3-8}
 &  & \multirow{2}{*}{$1$} & raw & 6.704 & 4.874 & 3.389 & 1.959\tabularnewline
 &  &  & adj & 6.545 & 4.137 & 2.657 & 1.587\tabularnewline
\cline{3-8}
 &  & \multirow{2}{*}{$\star$} & raw & 4.997 & 3.488 & 2.445 & 1.550\tabularnewline
 &  &  & adj & 5.595 & 3.672 & 2.492 & 1.562\tabularnewline
\cline{2-8}
 & \multirow{2}{*}{Oracle} &  & raw & 5.112 & 3.581 & 2.507 & 1.566\tabularnewline
 &  &  & adj & 5.316 & 3.595 & 2.492 & 1.562\tabularnewline
\cline{2-8}
 & BETEL &  &  & 4.354 & 3.210 & 2.343 & 1.513\tabularnewline
\hline 
\end{tabular}
\end{table}

\clearpage{}

\begin{table}
\caption{Quasi-posterior summaries for $\theta_{j}$ across CCQB calibration
stages}

\label{tab:wheeze-theta}

\medskip{}

\centering{}%
\begin{tabular}{clccc}
\hline 
Param. &  & \multicolumn{3}{c}{\begin{cellvarwidth}[t]
\centering
Quasi-posterior mean

{[}$95\%$ interval{]}
\end{cellvarwidth}}\tabularnewline
\cline{3-5}
 &  & $s=0$ & $s=1$ & $s=\star$\tabularnewline
\hline 
\multirow{3}{*}{$\theta_{1}$} &  & -2.039 & -1.917 & -1.918\tabularnewline
 & raw & {[}-2.968, -1.447{]} & {[}-2.156, -1.703{]} & {[}-2.156, -1.705{]}\tabularnewline
 & adj & {[}-2.299, -1.778{]} & {[}-2.151, -1.683{]} & {[}-2.153, -1.683{]}\tabularnewline
\hline 
\multirow{3}{*}{$\theta_{2}$} &  & -0.141 & -0.135 & -0.136\tabularnewline
 & raw & {[}-0.699, 0.352{]} & {[}-0.241, -0.030{]} & {[}-0.242, -0.032{]}\tabularnewline
 & adj & {[}-0.269, -0.013{]} & {[}-0.250, -0.020{]} & {[}-0.251, -0.021{]}\tabularnewline
\hline 
\multirow{3}{*}{$\theta_{3}$} &  & 0.081 & 0.219 & 0.222\tabularnewline
 & raw & {[}-0.730, 0.829{]} & {[}-0.142, 0.576{]} & {[}-0.149, 0.569{]}\tabularnewline
 & adj & {[}-0.381, 0.542{]} & {[}-0.165, 0.604{]} & {[}-0.162, 0.607{]}\tabularnewline
\hline 
\multirow{3}{*}{$\theta_{4}$} &  & 0.001 & 0.046 & 0.048\tabularnewline
 & raw & {[}-0.223, 0.225{]} & {[}-0.095, 0.186{]} & {[}-0.093, 0.188{]}\tabularnewline
 & adj & {[}-0.210, 0.212{]} & {[}-0.135, 0.227{]} & {[}-0.133, 0.228{]}\tabularnewline
\hline 
\end{tabular}
\end{table}

\clearpage{}

\begin{table}
\caption{Quasi-posterior summaries for $\operatorname{OR}(a)$ across CCQB
calibration stages}

\label{tab:wheeze-or}

\medskip{}

\centering{}%
\begin{tabular}{clccc}
\hline 
Age &  & \multicolumn{3}{c}{\begin{cellvarwidth}[t]
\centering
Quasi-posterior mean

{[}$95\%$ interval{]}
\end{cellvarwidth}}\tabularnewline
\cline{3-5}
 &  & $s=0$ & $s=1$ & $s=\star$\tabularnewline
\hline 
\multirow{3}{*}{7} &  & 1.191 & 1.16 & 1.158\tabularnewline
 & raw & {[}0.441, 2.498{]} & {[}0.762, 1.676{]} & {[}0.760, 1.670{]}\tabularnewline
 & adj & {[}0.674, 1.707{]} & {[}0.672, 1.648{]} & {[}0.671, 1.646{]}\tabularnewline
\hline 
\multirow{3}{*}{8} &  & 1.173 & 1.209 & 1.210\tabularnewline
 & raw & {[}0.474, 2.317{]} & {[}0.833, 1.684{]} & {[}0.833, 1.675{]}\tabularnewline
 & adj & {[}0.719, 1.627{]} & {[}0.774, 1.643{]} & {[}0.775, 1.645{]}\tabularnewline
\hline 
\multirow{3}{*}{9} &  & 1.171 & 1.266 & 1.270\tabularnewline
 & raw & {[}0.482, 2.292{]} & {[}0.868, 1.779{]} & {[}0.861, 1.767{]}\tabularnewline
 & adj & {[}0.670, 1.671{]} & {[}0.788, 1.745{]} & {[}0.790, 1.750{]}\tabularnewline
\hline 
\multirow{3}{*}{10} &  & 1.184 & 1.333 & 1.340\tabularnewline
 & raw & {[}0.461, 2.417{]} & {[}0.855, 1.963{]} & {[}0.854, 1.960{]}\tabularnewline
 & adj & {[}0.552, 1.816{]} & {[}0.712, 1.955{]} & {[}0.715, 1.965{]}\tabularnewline
\hline 
\end{tabular}
\end{table}

\clearpage{}

\begin{table}
\caption{Quasi-posterior summaries for $\theta_{2}$ across CCQB calibration
stages}

\label{tab:wheeze-ablation}

\medskip{}

\centering{}%
\begin{tabular}{clccc}
\hline 
$\varphi$ &  & \multicolumn{3}{c}{\begin{cellvarwidth}[t]
\centering
Quasi-posterior mean

{[}$95\%$ interval{]}
\end{cellvarwidth}}\tabularnewline
\cline{3-5}
 &  & $s=0$ & $s=1$ & $s=\star$\tabularnewline
\hline 
\multirow{3}{*}{0.2} &  & -0.147 & -0.135 & -0.135\tabularnewline
 & raw & {[}-0.760, 0.377{]} & {[}-0.241, -0.033{]} & {[}-0.242, -0.031{]}\tabularnewline
 & adj & {[}-0.276, -0.017{]} & {[}-0.250, -0.020{]} & {[}-0.251, -0.020{]}\tabularnewline
\hline 
\multirow{3}{*}{1} &  & -0.141 & -0.135 & -0.136\tabularnewline
 & raw & {[}-0.699, 0.352{]} & {[}-0.241, -0.030{]} & {[}-0.242, -0.032{]}\tabularnewline
 & adj & {[}-0.269, -0.013{]} & {[}-0.250, -0.020{]} & {[}-0.251, -0.021{]}\tabularnewline
\hline 
\multirow{3}{*}{100} &  & -0.064 & -0.135 & -0.135\tabularnewline
 & raw & {[}-0.384, 0.327{]} & {[}-0.239, -0.033{]} & {[}-0.240, -0.032{]}\tabularnewline
 & adj & {[}-0.182, 0.054{]} & {[}-0.250, -0.020{]} & {[}-0.251, -0.020{]}\tabularnewline
\hline 
\end{tabular}
\end{table}

%\begin{comment}
\includepdf[pages=-]{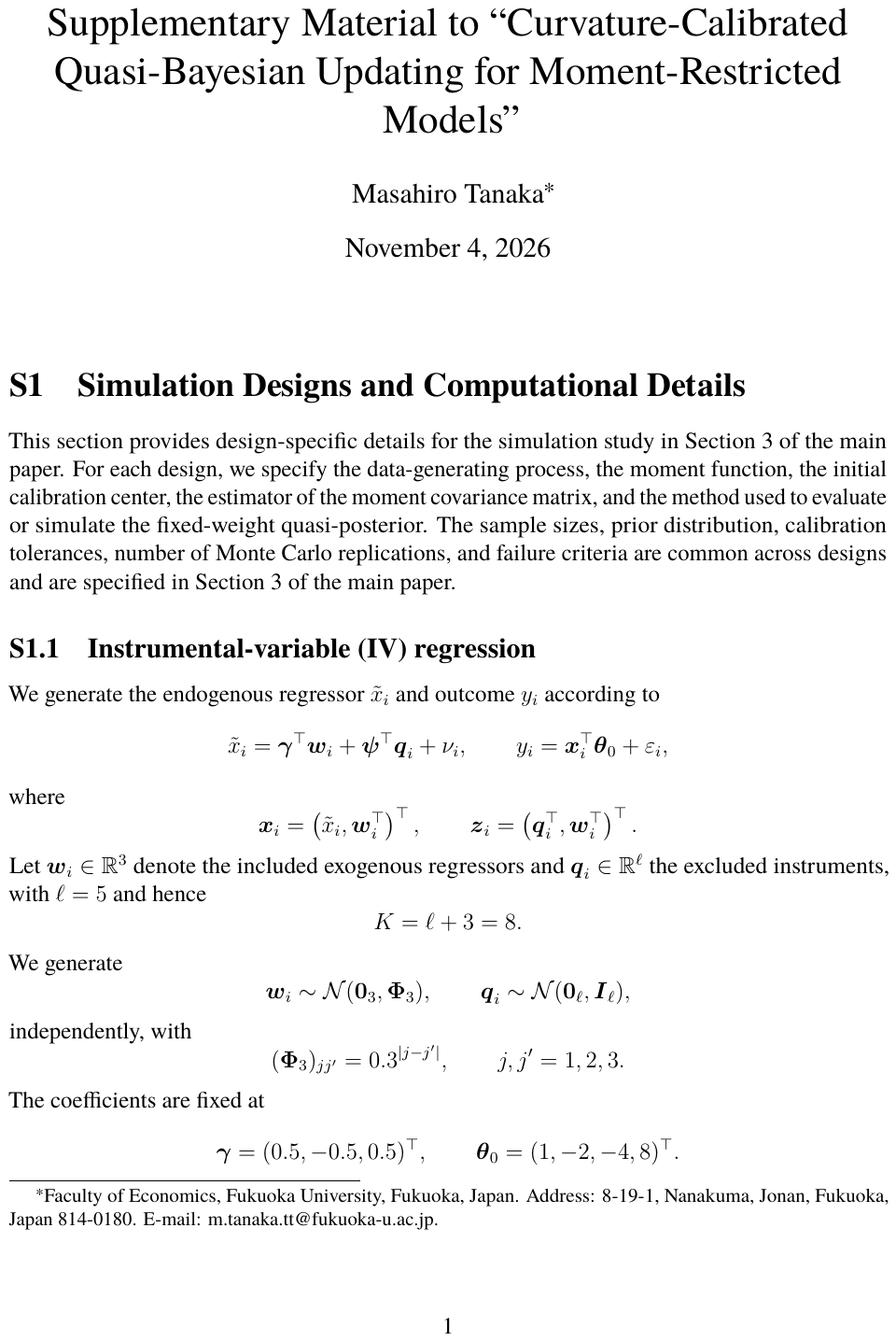}
%\end{comment}

\end{document}